\pdfoutput=1
\RequirePackage{ifpdf}
\ifpdf 
\documentclass[pdftex]{sigma}
\else
\documentclass{sigma}
\fi

\numberwithin{equation}{section}

\newtheorem{Theorem}{Theorem}[section]
\newtheorem*{Theorem*}{Theorem}

\theoremstyle{definition}

\newtheorem{Example}[Theorem]{Example}

\usepackage{bm}
\usepackage{mathstyle}

\newcommand{\CA}{{\cal A}}
\newcommand{\CB}{{\cal B}}
\newcommand{\CC}{{\cal C}}

\newcommand{\CH}{{\cal H}}

\newcommand{\CJ}{{\cal J}}
\newcommand{\CK}{{\cal K}}
\newcommand{\CL}{{\cal L}}
\newcommand{\CM}{{\cal M}}

\newcommand{\CO}{{\cal O}}

\newcommand{\CS}{{\cal S}}

\newcommand{\CW}{{\cal W}}

\def\IZ{{\mathbb Z}}
\def\IR{{\mathbb R}}
\def\IH{{\mathbb H}}
\def\IC{{\mathbb C}}
\def\IQ{{\mathbb Q}}
\def\IN{{\mathbb N}}

\def\IT{{\mathbb T}}
\def\IS{{\mathbb S}}

\newcommand\TT{{\mathbb T}}

\def\FE{\bm{4}_1}
\def\TT{\bm{5}_2}
\def\e{\epsilon}
\def\u{\upsilon}

\newcommand{\tr}{\operatorname{Tr}}
\newcommand{\re}{{\rm e}}

\newcommand{\ri}{{\rm i}}
\newcommand{\rd}{{\rm d}}

\renewcommand{\d}{\partial}
\def\Hom{\operatorname{Hom}}

\newcommand{\mC}{\mathsf{C}}

\def\tq{{\tilde{q}}}
\def\tD{{\tilde{D}}}

\def\tq{{\tilde{q}}}

\newcommand{\mb}{{\mathsf{b}}}

\newcommand{\be}{\begin{equation}}
\newcommand{\ee}{\end{equation}}
\newcommand{\ba}{\begin{aligned}}
\newcommand{\ea}{\end{aligned}}

\begin{document}
\allowdisplaybreaks

\newcommand{\arXivNumber}{2312.00624}

\renewcommand{\PaperNumber}{002}

\FirstPageHeading

\ShortArticleName{On the Structure of Wave Functions in Complex Chern--Simons Theory}

\ArticleName{On the Structure of Wave Functions\\ in Complex Chern--Simons Theory}

\Author{Marcos MARI\~NO~$^{\rm a}$ and Claudia RELLA~$^{\rm b}$}

\AuthorNameForHeading{M.~Mari\~no and C.~Rella}

\Address{$^{\rm a)}$~D\'epartement de Physique Th\'eorique and Section de Math\'ematiques,\\
\hphantom{$^{\rm a)}$}~Universit\'e de Gen\`eve, CH-1211 Gen\`eve, Switzerland}
\EmailD{\mail{marcos.marino@unige.ch}}

\Address{$^{\rm b)}$~Institut des Hautes \'Etudes Scientifiques, 91440 Bures-sur-Yvette, France}
\EmailD{\mail{rella@ihes.fr}}

\ArticleDates{Received February 10, 2025, in final form December 19, 2025; Published online January 06, 2026}

\Abstract{We study the structure of wave functions in complex Chern--Simons theory on the complement of a hyperbolic knot, emphasizing the similarities with the topological string/spectral theory correspondence. We first conjecture a hidden integrality structure in the holomorphic blocks and show that this structure guarantees the cancellation of potential singularities in the full non-perturbative wave function at rational values of the coupling constant. We then develop various techniques to determine the wave function at such rational points. Finally, we illustrate our conjectures and obtain explicit results in the examples of the figure-eight and three-twist knots. In the case of the figure-eight knot, we also perform a direct evaluation of the state integral in the rational case and observe that the resulting wave function has the features of the ground state for a quantum mirror curve.}

\Keywords{complex Chern--Simons theory; hyperbolic knots; $A$-polynomial; wave functions; holomorphic blocks; integrality structure}

\Classification{57K31; 81T45; 81Q05}

\section{Introduction} \label{sec: intro}
Chern--Simons (CS) theory with a complex gauge group has been an excellent laboratory for studying various aspects of quantum field theory (QFT) since it is essentially exactly solvable.
In recent years, building on previous work by physicists and mathematicians, perturbative and non-perturbative methods have been introduced, making it possible
to calculate various observables, and many beautiful and interesting results have been obtained.
For example, for the complements of hyperbolic knots in the three-sphere, the wave function of the theory with gauge group $\mathrm{SL}(2,\IC)$ has been defined rigorously~\cite{ak}, inspired in part by physics developments~\cite{dglz,hikami07}.
This wave function satisfies, in addition, a difference equation~\cite{amalusa}, which can be determined by an appropriate quantization of the classical $A$-polynomial of the knot~\cite{stavros-aj}, as expected from physics arguments~\cite{gukov}.

In this paper, we will further study the wave functions for complements of hyperbolic knots in CS theory with gauge group $\mathrm{SL}(2,\IC)$.
As we will show, they share many structural similarities with the wave functions occurring in topological string theory and, more precisely, in the context of the so-called topological string/spectral theory (TS/ST) correspondence~\cite{cgm,ghm, mmrev,wzh} and its open string version~\cite{fgrassi,mz-wv,mz-wv2,szabolcs}. Indeed, it has been found in both cases that the WKB expansion in $\hbar$ of the perturbative wave function of the topological string can be resummed into a $q$-series,\footnote{This is a resummation of a {\it convergent} series, so it does not involve the more sophisticated summability techniques that apply to divergent series, like the Borel--Laplace resummation.} where $q= \re^{\ri \hbar}$. However, this $q$-series displays singularities at all points of the form%
\begin{gather} \label{hsingular}
\hbar \in 2 \pi \IQ .
\end{gather}
Since these singularities are not present in the non-perturbative definition of the wave function,
they are an artifact of the resummed WKB expansion. This feature, in turn, requires the presence of a
non-perturbative sector that cancels the singularities and leads to a finite answer for the full wave function.
The cancellation of singularities at the values of $\hbar$ in equation~\eqref{hsingular} is a defining attribute of Faddeev's quantum dilogarithm~\cite{faddeev}, which can be regarded as a simple example of a wave function in complex CS theory.
A similar cancellation mechanism has appeared in a more complicated context in ABJM theory, where it is sometimes referred to as the
HMO mechanism~\cite{hmo2}, and is one of the facets of the non-perturbative proposal for topological string theory put forward in~\cite{cgm,ghm,wzh}.
Moreover, the concrete realization of the cancellation of singularities in topological string theory is a consequence of the integrality structure of the topological string amplitudes.
In the closed string case, one has the Gopakumar--Vafa integrality~\cite{gv} and its refinement~\cite{hiv}.
In the open string case, as shown in~\cite{kpamir,mz-wv,szabolcs}, one needs an integrality
property akin to the one found in~\cite{nawata,lmv,ov}. In this paper, we conjecture an integrality structure for
the resummed WKB expansion of the wave functions of complex CS theory and support our claims with explicit evidence obtained in examples.
This integrality structure, indeed, guarantees the cancellation of singularities.
A similar integrality property has been found in~\cite{ekholm}, and it would be interesting to clarify its relationship with our results.\looseness=1

Our integrality conjecture characterizes the so-called holomorphic blocks of the wave functions~\cite{bdp}, and the corresponding integer invariants can be calculated from these blocks when they are explicitly known.
Analogously to the case of topological string theory, at the special values of $\hbar$ in equation~\eqref{hsingular}, which we refer to as rational points, the generic formula for the exact wave function has an apparent singularity. However, after the cancellation mechanism takes place, one typically finds a relatively simpler expression. It is an interesting task to determine this expression for
general rational numbers, as it was done for Faddeev's quantum dilogarithm in~\cite{garou-kas}.
In this paper, we develop various independent techniques to achieve this goal, following ideas proposed in the context of the open TS/ST correspondence~\cite{butterfly,hsx,szabolcs}.
To begin with, one can use the underlying integrality structure to derive an explicit answer at rational values of $\hbar$ in terms of the newly introduced integer invariants.
Second, one can start directly from the AJ equation for the wave function and specialize it to rational values, where one finds a quasi-periodic structure similar to the one appearing in spectral problems on lattices.
Finally, one can directly evaluate the Andersen--Kashaev state integral at rational points using the techniques of~\cite{garou-kas}.
We apply these three distinct methods and explicitly show that they lead to the same results in examples.

The paper is organized as follows.
In Section~\ref{sec: theory}, we review the necessary background notions on complex CS theory and the AJ conjecture.
In Section~\ref{sec: main}, we present most of our results. In particular, we state the integrality conjecture for the WKB resummed wave function and show how it implies the cancellation of singularities at rational values of $\hbar$.
We also provide two different techniques for evaluating the wave function at rational points.
In Section~\ref{sec: examples}, we illustrate these results by performing explicit computations in the examples of the figure-eight and three-twist knots.
In Section~\ref{sec: stateintegral}, we evaluate directly the state integral in the rational case for the figure-eight knot.
Finally, we conclude and list some open problems in Section~\ref{sec-conclusions}.
In the three appendices, we provide additional details on the calculations we perform in Section~\ref{sec: main} and recall some useful properties of Faddeev's quantum dilogarithm.

\section[Complex Chern--Simons theory and the A-polynomial]{Complex Chern--Simons theory and the $\boldsymbol{ A}$-polynomial} \label{sec: theory}
In this section, we review the fundamental aspects of complex CS theory on a closed three-manifold and the construction of the classical and quantum $A$-polynomials of a hyperbolic knot.
The physical understanding of the connection to $A$-polynomials was
developed in~\cite{gukov}, and an excellent summary can be found in~\cite{dimofte-rev}.
The AJ conjecture, which is at the basis of many of our computations, was proposed in~\cite{stavros-aj}.
We pay special attention to the case of CS theory with~$\mathrm{SL}(2,\IC)$ gauge group on the complement of a hyperbolic knot in the three-sphere and introduce the two benchmark examples that we will consider later in this work, that is, the figure-eight and three-twist knots.

\subsection[Classical and quantum A-polynomials]{Classical and quantum $\boldsymbol{ A}$-polynomials} \label{sec: Apoly}
The classical action of CS theory with complex gauge group $G_{\IC}$ can be written as~\cite{witten91}
\be \label{CS-action}
S = \frac{t}{8 \pi} \int_M \tr\left(\CA \wedge \rd\CA + \frac{2}{3} \CA \wedge \CA \wedge \CA \right) + \frac{t'}{8 \pi} \int_M \tr\left(\bar{\CA} \wedge \rd\bar{\CA} + \frac{2}{3} \bar{\CA} \wedge \bar{\CA} \wedge \bar{\CA} \right) ,
\ee
where $t$, $t'$ are complex parameters, $M$ is the underlying closed three-manifold with boundary~${\Sigma = \d M}$, and
$\CA$ is the complex gauge field, which is a one-form on $M$ taking values in the Lie algebra $\mathfrak{g}_{\IC}$. Here, $\bar{\CA}$ denotes the complex conjugate of $\CA$.
The coefficients $t$, $t'$ can be written as~${t=k+\ri s}$ and $t'=k-\ri s$. Here, quantization of the theory implies that the level $k$ is an integer, while unitarity requires $s$ to be either real or purely imaginary, although we will not use this in our discussion.
We also introduce the complex coupling constant $\mb^2$ defined by
\begin{gather}
\mb^2= \frac{t'}{t} .
\end{gather}

A classical solution on $M$ is identified with a gauge equivalence class of flat $G_{\IC}$-connections on $M$, which are gauge fields $\CA$ satisfying the classical Euler--Lagrange equations
\begin{gather}
\rd \CA + \CA \wedge \CA = \rd \bar{\CA} + \bar{\CA} \wedge \bar{\CA} = 0 .
\end{gather}
Because a flat $G_{\IC}$-connection on $M$ is determined by a homomorphism
\be \label{rho-hom}
\rho \colon\ \pi_1(M) \rightarrow G_{\IC} ,
\ee
the moduli space of classical solutions on $M$ is thus
\be \label{classical_sol}
\CM_{{\rm flat}} (G_{\IC}, M) = \Hom (\pi_1(M), G_{\IC} ) / {\sim},
\ee
where $\pi_1$ denotes the fundamental group and $\sim$ is conjugation by elements of the gauge group~$G_{\IC}$.
The classical phase space of the theory is instead given by the moduli space of flat $G_{\IC}$-connections on the closed Riemann surface $\Sigma$ modulo gauge transformations, that is,
\be \label{phase_space}
\CM_{{\rm flat}} (G_{\IC}, \Sigma) = \Hom (\pi_1(\Sigma), G_{\IC}) / {\sim},
\ee
which comes naturally equipped with the symplectic two-form
\be \label{omega}
\omega = \frac{t}{8 \pi} \int_{\Sigma} \tr (\delta \CA \wedge \delta \CA ) + \frac{t'}{8 \pi} \int_{\Sigma} \tr \bigl(\delta \bar{\CA} \wedge \delta \bar{\CA} \bigr) .
\ee
Recall that there is a natural map
\begin{gather}
\iota \colon\ \CM_{{\rm flat}} (G_{\IC}, M) \rightarrow \CM_{{\rm flat}} (G_{\IC}, \Sigma)
\end{gather}
induced by the inclusion $\pi_1(\Sigma) \hookrightarrow \pi_1(M)$. Indeed, the image of the moduli space of classical solutions in equation~\eqref{classical_sol} under this map is a Lagrangian submanifold of the classical phase space.
Namely,
\begin{gather}
\CL = \iota(\CM_{{\rm flat}} (G_{\IC}, M) )\subset \CM_{{\rm flat}} (G_{\IC}, \Sigma)
\end{gather}
is Lagrangian with respect to the symplectic structure in equation~\eqref{omega}.

In this paper, we will focus on a three-manifold $M$ obtained as the complement of a hyperbolic knot\footnote{A three-manifold $M$ is hyperbolic if there is a discrete faithful representation of its fundamental group $\pi_1(M)$ into the group of orientation-preserving isometries of $\IH^3$, that is, the group $\mathrm{Isom}^+\bigl(\IH^3\bigr) \cong \mathrm{PSL}(2, \IC)$. A hyperbolic knot is a knot whose complement in the three-sphere is hyperbolic.} $\CK$ in the three-sphere $\IS^3$, that is,
\be\label{MK}
M = \IS^3 \backslash \CK ,
\ee
which has a single toral boundary $\Sigma = \IT^2$. Its moduli space of flat $G_{\IC}$-connections identifies a~complex Lagrangian submanifold $\CL$ of the full phase space
\be \label{phase_spaceK}
\CM_{{\rm flat}} \bigl(G_{\IC}, \IT^2\bigr) = ( \IT_{\IC} \times \IT_{\IC} ) / \CW ,
\ee
where $\IT_{\IC}$ is the maximal toral subgroup of $G_{\IC}$ and $\CW$ is the Weyl group.
Let us denote by~${\bm{P}=(P_1, \dots, P_r)}$ and $\bm{X}=(X_1, \dots, X_r)$, where $r$ is the rank of the gauge group, the complex variables parametrizing each copy of the maximal torus $\IT_{\IC}$ in equation~\eqref{phase_spaceK}, which are defined modulo the action of the Weyl group.\footnote{$\bm{P}$ and $\bm{X}$ can be interpreted as the vectors of eigenvalues of the holonomies of the flat gauge connections on the boundary torus over its two basic one-cycles.}
It follows that the irreducible components of $\CL$ are described by $\CW$-invariant polynomial equations
\be \label{Apoly_eqs}
A_{\CK, i} (\bm{P}, \bm{X}) = 0 , \qquad i=1, \dots, r ,
\ee
where the polynomials $A_{\CK, i} (\bm{P}, \bm{X})$ have coefficients in $\IZ$~\cite{cooper}.

Quantizing the classical phase space in equation~\eqref{phase_space} with its symplectic structure in equation~\eqref{omega} produces an infinite-dimensional Hilbert space $\CH_{\Sigma}$, and the Feynman path integral over the manifold $M$ leads to a state $|M\rangle \in \CH_{\Sigma}$. In the case of a hyperbolic knot complement, the polynomials $A_{\CK, i}(\bm{P}, \bm{X})$ in equation~\eqref{Apoly_eqs} are expected to produce quantum operators~\smash{$\hat{A}_{\CK, i}\bigl(\hat{\bm{P}}, \hat{\bm{X}},q\bigr)$}, $i=1, \dots, r$, acting on $\CH_{\Sigma}$, which annihilate the state $|M\rangle$. Here, the complex variables $\bm{P}$, $\bm{X}$ are promoted to operators $\hat{\bm{P}}$, $\hat{\bm{X}}$ satisfying the commutation relations
\begin{gather}
\hat{P}_i \hat{X}_j = q^{\delta_{ij}} \hat{X}_j \hat{P}_i , \qquad i,j=1, \dots, r ,
\end{gather}
where $\delta_{ij}$ is the Kronecker delta, we have introduced $q= \re^{\ri \hbar}$, and
\begin{gather}
\hbar = 2 \pi \mb^2
\end{gather}
is the complex coupling parameter playing the role of Planck's constant.
We will sometimes denote $\tau= \mb^2$.
Note that the quantization of the theory depends on the level $k$ through the parameter $\mb^2$, and, in this work, we will restrict ourselves to the case $k=1$.
See, e.g.,~\cite{dimofte-levelk}.

The classical constraints in equation~\eqref{Apoly_eqs} become Schr\"odinger-like operator equations in the quantum theory. Namely,
\begin{gather} \label{quantumApoly}
\hat{A}_{\CK, i}\bigl(\hat{\bm{P}}, \hat{\bm{X}},q\bigr) \chi_{\CK}(\hbar) = 0 , \qquad i=1, \dots, r ,
\end{gather}
where $\chi_\CK(\hbar)$ is the partition function associated with the manifold $M$ in equation~\eqref{MK}. Following the conventions of~\cite{ggm2}, we define the continuous complex variables $\bm{u}=(u_1, \dots, u_r)$ such that
\begin{gather}
X_i = \re^{2 \pi \mb u_i} , \qquad i=1, \dots, r .
\end{gather}
The corresponding quantum operators $\hat{u}_i$ determine a complete basis of states $|u\rangle \in \CH_{\Sigma}$ on which they act by multiplication.
Therefore, the partition function on $M$ can be regarded as a~wave function in $u$-space, which we represent as
\begin{gather}
\chi_{\CK}(\bm{u}; \hbar) .
\end{gather}
We stress that the quantization of the $A$-polynomials in equation~\eqref{Apoly_eqs} is not obtained simply by promoting $\bm{P}$, $\bm{X}$ to their operator counterparts.
On top of ordering issues, the resulting operators have a non-trivial dependence on $\hbar$ through $q$.
It is fair to say that the spectral theory of these operators is not entirely understood, and a deeper grasp of this issue might clarify the corresponding non-trivial quantization problem.
A concrete way of constructing the quantum $A$-polynomials is, for example, to use the original AJ conjecture of~\cite{stavros-aj}.

The wave function $\chi_{\CK}(\bm{u}; \hbar)$ can be computed perturbatively in a saddle-point approximation where the saddle, or classical solution of the Euler--Lagrange equations, is described by the group homomorphism in equation~\eqref{rho-hom}.
Equivalently, the saddle points can be identified with the different classical solutions $\bm{P}^{(\alpha)}$ to equation~\eqref{Apoly_eqs} for fixed $\bm{X}$, which we mark with the discrete label $\alpha$. We recall that the branches of the polynomials in equation~\eqref{Apoly_eqs} come in conjugate pairs due to the symmetry of the theory under conjugation. Thus, each flat connection $\CA^{(\rho)}$, labeled by $\rho \in \Hom (\pi_1(M), G_{\IC})$, has a conjugate flat connection $\CA^{(\bar{\rho})}$, corresponding to the conjugate homomorphism $\bar{\rho}$.
Consequently, we will denote by $\bar{\alpha}$ the branch conjugate to $\alpha$.
Note that there is always an abelian branch $\alpha = \mathrm{abel}$ described by the equations $P_1=\dots=P_r=1$, which is self-conjugate, and a geometric branch\footnote{On the geometric branch, $\bm{u}$ can be thought of as parametrizing the quantum deformations of the complex hyperbolic structure of $M$.} $\alpha = {\rm geom}$ containing the discrete faithful representation of $\pi_1(M)$ into $\mathrm{PSL}(2, \IC)$, which has a distinct conjugate $\alpha = {\rm conj}$.
In the rest of this work, we will only consider non-abelian branches. Therefore, we simplify the factor corresponding to the abelian classical solution from the polynomials in equation~\eqref{Apoly_eqs} and proceed to quantize the simplified form.

When computed in the saddle-point approximation, that is, using the WKB method, around the classical solution labeled by $\alpha$, the perturbative wave function is denoted by \smash{$\chi_{\CK}^{(\alpha), {\rm WKB}}(\bm{u};\hbar)$} and given by an asymptotic series in $\hbar$ of the form~\cite{dglz}
\be \label{WKB}
 \chi_{\CK}^{(\alpha), {\rm WKB}}(\bm{u};\hbar) = \exp \left( \frac{1}{\hbar} S_0^{(\alpha)}(\bm{u}) - \frac{\delta^{(\alpha)}}{2} \log \hbar + \sum_{n=0}^{\infty} S_{n+1}^{(\alpha)}(\bm{u}) \hbar^n \right) ,
\ee
where the leading-order coefficient $S_0^{(\alpha)}(\bm{u})$ is determined by the classical CS functional in equation~\eqref{CS-action} up to integer multiples of $2 \pi \ri \bm{u}$, the next-to-leading-order coefficient $\delta^{(\alpha)}$ is an integer that vanishes when $\alpha \ne \mathrm{abel}$, and the higher-order coefficients \smash{$S_n^{(\alpha)}(\bm{u})$}, $n \ge 1$, are obtained, in principle, by summing the contributions of $n$-loops Feynman diagrams systematically.

\subsection[The case of SL(2,IC) gauge group]{The case of $\boldsymbol{\mathrm{SL}(2,\IC)}$ gauge group} \label{sec: sl2c}
In this work, we take $G_{\IC} = \mathrm{SL}(2,\IC)$. Therefore, we have simply $r=1$ and
\begin{gather}
A_{\CK}(P,X)=A_{\CK, 1}(P_1,X_1)
\end{gather}
is the classical $A$-polynomial of the knot~\cite{cooper}.
The operators $\hat{P}$, $\hat{X}$ act on $u$-states as
\begin{gather}
\hat{P} |u\rangle = |u+ \ri \mb \rangle , \qquad \hat{X} |u\rangle = X |u\rangle .
\end{gather}
Equivalently, $\hat{P}$ is the operator that shifts $X$ into $q X$, and $\hat{X}$ is the ordinary multiplication.
Building on previous results~\cite{dglz,hikami01, hikami07}, the exact, non-perturbative partition function $\chi_{\CK}(u; \hbar)$ has been constructed from an ideal triangulation of $M$ and identified with a finite-dimensional integral whose integrand is a product of Faddeev’s quantum dilogarithm functions~\cite{faddeev}, known as the state-integral invariant, or Andersen--Kashaev invariant~\cite{ak, ak2, dimofte-rev}. This is a holomorphic function of $\hbar \in \IC' = \IC \backslash (-\infty, 0 ]$ and $u \in \IC$. It has been explicitly verified in the examples of the figure-eight and three-twist knots and conjectured in general~\cite{amalusa} that it obeys the $q$-difference equation encoded in the quantum $A$-polynomial of the knot, that is,
\be \label{AJ-eq}
\hat{A}_{\CK}\bigl(\hat{P}, \hat{X},q\bigr) \chi_{\CK}(u;\hbar) = 0 .
\ee

Recall that $S$-duality exchanges $\mb$ with $1/\mb$. Therefore, the $S$-dual images of $\hbar$ and $q$ are
\begin{gather}
\hbar_D={4 \pi^2 \over \hbar} = \frac{2 \pi}{\tau} , \qquad q_D= \re^{\ri \hbar_D} = \re^{\frac{2 \pi \ri}{\tau}}= \tilde q^{-1} ,
\end{gather}
where we have introduced \smash{$\tq = \re^{-{2 \pi \ri \over \tau}}$}, while the variable $u$ is invariant under the action of $S$-duality and $X$ transforms into
\begin{gather}
X_D = \re^{\frac{2 \pi u}{\mb}} = X^{2 \pi \over \hbar} .
\end{gather}
Remarkably, the wave function $\chi_{\CK}(u;\hbar)$ is also invariant under this transformation, that is,
\begin{gather}
\chi_{\CK}(u;\hbar)=\chi_{\CK}(u;\hbar_D) ,
\end{gather}
and, in particular, it satisfies the $S$-dual quantum $A$-polynomial equation\footnote{
Since the wave function simultaneously obeys distinct $q$- and $q_D$-difference equations, the solution space of equations~\eqref{AJ-eq} and~\eqref{AJ-eq-dual} may be naturally regarded as a bimodule over the difference operator algebras associated with the quantum $A$-polynomial and its $S$-dual image. We will not, however, pursue this perspective further here.
}
\be \label{AJ-eq-dual}
\hat A_\CK\bigl(\hat{P}_D, \hat{X}_D, q_D\bigr)\chi_{\CK}(u;\hbar)=0 ,
\ee
where $\hat{P}_D$ is the quantum operator acting on $u$-states by shifting $u$ into $u + \ri/\mb$, or, equivalently, transforming $X_D$ into $q_D X_D$, and $\hat{X}_D$ acts by multiplication.

\begin{figure}[htb!]
\center
 \includegraphics[width=0.25\textwidth]{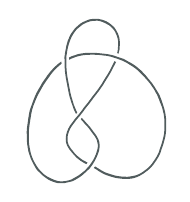} \qquad \qquad
 \includegraphics[width=0.25\textwidth]{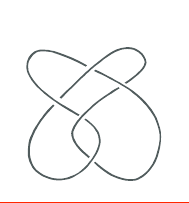}
 \caption{The hyperbolic knots known as the figure-eight knot ($\FE$), on the left side, and the three-twist knot ($\TT$), on the right side.}
 \label{fig: simple_knots}
\end{figure}

Throughout this work, we examine the two simplest examples of hyperbolic knots, which allow us to perform explicit and detailed calculations to support and give insight into our claims. These are the figure-eight and three-twist knots.

\begin{Example} \label{41example}
The figure-eight knot, denoted as $\FE$, is shown on the left side of Figure~\ref{fig: simple_knots}.
Its classical $A$-polynomial can be written in symmetrized form as
\begin{gather}
A_{\FE}(P, X) = P^{-1} + P -\bigl( X^2+ X^{-2} -X -X^{-1} -2\bigr) .
\end{gather}
The corresponding polynomial equation $A_{\FE}(P, X)=0$ has only two distinct non-abelian branch\-es.
Explicitly, the geometric and conjugate branches are labeled as
\begin{gather}
\alpha = {\rm geom}, {\rm conj} ,
\end{gather}
and they are described by
\be \label{classicP}
P^{({\rm geom}, {\rm conj})} = \frac{T(X) \pm \bigl(1-X^2\bigr) \sqrt{\Delta(X)} }{2 X^2} ,
\ee
where we have introduced
\begin{subequations}
\begin{gather}
T(X) = 1 - X - 2 X^2 - X^3 + X^4 , \label{classicT} \\
\Delta(X) = 1 - 2 X - X^2 - 2 X^3 + X^4 . \label{classicD}
\end{gather}
\end{subequations}
Building on the results and conventions of~\cite{ggm2}, the descendant $\hat A$-operator is given by
\begin{equation}
 \label{rec41x}
 \hat A_{\bf{4}_1} \bigl(\hat{P},\hat{X},q^m,q\bigr) = \sum_{j=0}^2 C_j\bigl(\hat{X},q^m,q\bigr) \hat{P}^j ,
\end{equation}
where $m$ takes values in $\IZ$ and the coefficient functions $C_j\bigl(\hat{X},q^m,q\bigr)$ are
\begin{subequations}
\begin{align}
 \label{rec41x0}
& C_0\bigl(\hat{X},q^m,q\bigr) = q^{2 + 3 m} \hat{X}^2 \bigl(-1 + q^{3 + m} \hat{X}^2\bigr) , \\
 \label{rec41x1}
& C_1\bigl(\hat{X},q^m,q\bigr) = -q^m \bigl(-1 + q^{2 + m} \hat{X}^2\bigr) \bigl(1 - q \hat{X} - q^{1 + m} \hat{X}^2\nonumber \\
 &\hphantom{C_1\bigl(\hat{X},q^m,q\bigr) =}{}
 - q^{3 + m} \hat{X}^2 - q^{3 + m} \hat{X}^3 + q^{4 + 2 m} \hat{X}^4\bigr) , \\
 \label{rec41x2}
& C_2\bigl(\hat{X},q^m,q\bigr) = q^2 \hat{X}^2 \bigl(-1 + q^{1 + m} \hat{X}^2\bigr) .
\end{align}
\end{subequations}
If we set $m=0$, symmetrize by multiplying by $\hat{P}^{-1}$, simplify a common factor $q^2 \hat{X}^2$, and perform the change of variable $\hat{X} \rightarrow q^{-1} \hat{X}$, we obtain the conventional
$\hat A$-operator of the figure-eight knot, that is,
\begin{equation}
 \label{41Ahat}
 \hat A_{\bf{4}_1} \bigl(\hat{P},\hat{X},q\bigr) = C_0\bigl(\hat{X},q\bigr) \hat{P}^{-1} + C_1\bigl(\hat{X},q\bigr) + C_2\bigl(\hat{X},q\bigr) \hat{P} ,
\end{equation}
where the coefficient functions $C_j\bigl(\hat{X},q\bigr)$ are
\begin{subequations}\label{41xC}
\begin{align}
 \label{41xC0}
& C_0\bigl(\hat{X},q\bigr) =q \hat{X}^2-1 ,\\
 \label{41xC1}
& C_1\bigl(\hat{X},q\bigr) =- \bigl(\hat{X}^2-1\bigr)\bigl( \hat{X}^2 + \hat{X}^{-2} - \hat{X} - \hat{X}^{-1} - q - q^{-1} \bigr) ,\\
 \label{41xC2}
& C_2\bigl(\hat{X},q\bigr) = q^{-1} \hat{X}^2-1 .
\end{align}
\end{subequations}
\end{Example}

\begin{Example} \label{52example}
The three-twist knot, denoted as $\TT$, is shown on the right side of Figure~\ref{fig: simple_knots}.
Its classical $A$-polynomial can be written in symmetrized form as
\begin{gather}
A_{\TT}(P, X) = P^{-1} -\bigl(X^{-2} -X^{-1} +2 X + 2 X^2 - X^3\bigr) \nonumber\\
\hphantom{A_{\TT}(P, X) =}{} - \bigl(X^{-1} - 2 - 2 X + X^3 - X^4\bigr) P - X^2 P^2 .
\end{gather}
The corresponding polynomial equation $A_{\TT}(P, X)=0$ has three distinct non-abelian branches: the geometric branch, its conjugate, and a third, self-conjugate branch. We label them by
\be
\alpha = {\rm geom}, {\rm conj}, {\rm self}.
\ee
Again, following~\cite{ggm2}, the descendant $\hat A$-operator is given by
\begin{equation}
 \label{rec52x}
 \hat A_{\bf{5}_2} \bigl(\hat{P},\hat{X},q^m,q\bigr) = \sum_{j=0}^3 C_j\bigl(\hat{X},q^m,q\bigr) \hat{P}^j ,
\end{equation}
where $m$ takes values in $\IZ$ and the coefficient functions $C_j\bigl(\hat{X},q^m,q\bigr)$ are
\begin{subequations}
\begin{align}
 \label{rec52x0}
 C_0\bigl(\hat{X},q^m,q\bigr) ={}
 &-q^{2 + m} \hat{X}^2 \bigl(1-q^2 \hat{X}\bigr)\bigl(1+q^2 \hat{X}\bigr)\bigl(1-q^5 \hat{X}^2\bigr) , \\
 C_1\bigl(\hat{X},q^m,q\bigr) ={}
 & \bigl(1-q \hat{X}\bigr)\bigl(1+q \hat{X}\bigr)\bigl(1-q^5 \hat{X}^2\bigr)\bigl(1-q\hat{X}-q\hat{X}^2-q^4\hat{X}^2+q^{2+m}\hat{X}^2 \nonumber\\
 &+q^{3+m}\hat{X}^2+q^2\hat{X}^3+q^5\hat{X}^3+q^5\hat{X}^4+q^{5+m}\hat{X}^4-q^6\hat{X}^5\bigr) , \label{rec52x1}\\
 C_2\bigl(\hat{X},q^m,q\bigr) ={}
 & q\hat{X}\bigl(1-q^2\hat{X}\bigr)\bigl(1+q^2\hat{X}\bigr)\bigl(1-q\hat{X}^2\bigr)\bigl(1-q^2\hat{X}-q^{2+m}\hat{X}-q^2\hat{X}^2-q^5\hat{X}^2 \nonumber\\
 &+q^4\hat{X}^3+q^7\hat{X}^3-q^{5+m}\hat{X}^3-q^{6+m}\hat{X}^3+q^7\hat{X}^4-q^9\hat{X}^5\bigr) , \label{rec52x2} \\
 C_3\bigl(\hat{X},q^m,q\bigr) ={}
 & q^{8+m}\hat{X}^4\bigl(1-q\hat{X}\bigr)\bigl(1+q\hat{X}\bigr)\bigl(1-q\hat{X}^2\bigr) . \label{rec52x3}
\end{align}
\end{subequations}
If we set $m=0$, symmetrize by multiplying by $\hat{P}^{-2}$, and perform the change of variable $\hat{X} \rightarrow q^{-1} \hat{X}$, we obtain the conventional
$\hat A$-operator of the three-twist knot, that is,
\be \label{52Ahat}
\hat A_{\bf{5}_2}\bigl(\hat{P},\hat{X},q\bigr) = C_0\bigl(\hat{X},q\bigr) \hat{P}^{-2} + C_1\bigl(\hat{X},q\bigr) \hat{P}^{-1} + C_2\bigl(\hat{X},q\bigr) + C_3\bigl(\hat{X},q\bigr) \hat{P} ,
\ee
where we have defined
\begin{subequations} \label{52xC}
\begin{align}
 \label{52xC0}
 C_0\bigl(\hat{X},q\bigr) ={}&-\hat{X}^2 \bigl(1-q^2 \hat{X}^2\bigr) \bigl(1-q^3 \hat{X}^2\bigr) , \\
 \label{52xC1}
 C_1\bigl(\hat{X},q\bigr) ={}& \bigl(1-\hat{X}^2\bigr) \bigl(1-q^3 \hat{X}^2\bigr)\bigl(q^2 \hat{X}^3-q^2 \hat{X}^2-q \hat{X}^5 \nonumber \\
 &+2 q \hat{X}^4+q^{-1}\hat{X}^3+q \hat{X}^2-q^{-1}\hat{X}^2+\hat{X}^2-\hat{X}+1\bigr) , \\
 \label{52xC2}
 C_2\bigl(\hat{X},q\bigr) ={}& \hat{X} \bigl(1-q^2 \hat{X}^2\bigr) \bigl(1-q^{-1}\hat{X}^2\bigr) \bigl(-q^4 \hat{X}^5+q^4 \hat{X}^3\nonumber \\
 &+q^3 \hat{X}^4-q^3 \hat{X}^3-q^3 \hat{X}^2-q^2 \hat{X}^3+q \hat{X}^3-2 q \hat{X}-\hat{X}^2+1\bigr) , \\
 \label{52xC3}
 C_3\bigl(\hat{X},q\bigr) ={}& q^4 \hat{X}^4 \bigl(1-\hat{X}^2\bigr)\bigl(1-q^{-1}\hat{X}^2\bigr) .
\end{align}
\end{subequations}
\end{Example}

\section{The structure of the wave function} \label{sec: main}

In this section, we study the non-perturbative wave function $\chi_{\CK}(u;\hbar)$ introduced in Section~\ref{sec: theory} by exploiting ideas from topological string theory. In particular, we propose new conjectures on the structure of this wave function, which are motivated by similar results in the open version of the TS/ST correspondence~\cite{kpamir, mz-wv,mz-wv2, szabolcs} and by the evidence presented in Section~\ref{sec: examples}.
In the following, we will not explicitly indicate the knot $\CK$ to avoid the cluttering of notation.

\subsection{General conjectures} \label{sec: conjectures}
Our {\it first claim} is that the WKB expansion in equation~\eqref{WKB} can be resummed at all orders in~$\hbar$ and order by order in
$X=\re^{2 \pi \mb u}$. Explicitly, for a fixed choice of $\alpha$, we have that
\be \label{WKB-conj}
\chi^{(\alpha), {\rm WKB}} (u; \hbar)=\exp \left( {\ri \over \hbar} s^{(\alpha)}_0(x) + s^{(\alpha)}_1(x) + \phi^{(\alpha), {\rm WKB}}(X;q) \right) ,
\ee
where $s^{(\alpha)}_{0,1}(x)$ are $\IR$-polynomials in the variable\footnote{The $S$-dual image of $x$ is $x_D = \log X_D = 2 \pi x /\hbar$.} $x= \log X$ and
\be \label{phi-struc}
\phi^{(\alpha), {\rm WKB}}(X;q) = \sum_{n \ge 1} a^{(\alpha)}_n(q) X^n + a^{(\alpha)}_{-n}(q) X^{-n} .
\ee
The coefficient functions $a^{(\alpha)}_{\pm n}(q)$ can be written as
\be \label{ahatf-def}
a^{(\alpha)}_{\pm n}(q)= {\hat{a}_{\pm n}^{(\alpha)} (q) \over n (q^n-1) q^{\sigma_{\alpha, \pm n}}} , \qquad n \in \IZ_{>0} ,
\ee
where $\hat{a}^{(\alpha)}_{\pm n} (q) \in \IZ [q]$ such that
\be
\hat{a}^{({\rm geom})}_{- n} (q) = 0 , \qquad n \in \IZ_{>0} ,
\ee
and $\sigma_{\alpha, m}$ is a function of $m \in \IZ_{\ne 0}$ that takes values in $\IN$, depends on the choice of classical branch $\alpha$, and satisfies
\be
\sigma_{{\rm geom}, m} = 0 , \qquad \sigma_{\alpha, - |m|} = 0 \qquad \forall \alpha .
\ee
In both examples of the figure-eight and three-twist knots studied in Section~\ref{sec: examples}, we have precise expressions for $\sigma_{\alpha, m}$ for all choices of $\alpha \ne {\rm geom}$, which are given in equations~\eqref{sigma41},~\eqref{sigma52}, and~\eqref{sigma52self}.
Moreover, our conjectural statements above simplify notably in the particular case of the geometric branch, whose function $\phi^{({\rm geom}), {\rm WKB}}(X;q)$ has a series expansion including only the positive powers of $X$ and whose coefficients \smash{$a^{({\rm geom})}_{n}(q)$} have a trivial factor of $q^{\sigma_{{\rm geom}, n}}=1$ in the denominator.\looseness=-1

Note that we use the same notation for the perturbative WKB series in equation~\eqref{WKB} and its resummation in terms of $q$ in equation~\eqref{WKB-conj}. However, in the rest of this paper, we will always refer to the resummed version of $\chi^{(\alpha), {\rm WKB}} (u; \hbar)$ as in equation~\eqref{WKB-conj}.
In fact, there is additional structure in the resummed perturbative wave function. Our {\it second claim} is that \smash{$\phi^{(\alpha), {\rm WKB}}(X;q)$} has an integrality/multicovering-type property.
Namely, for a fixed choice of $\alpha$, we have that
\be \label{multi-conj}
\phi^{(\alpha), {\rm WKB}}(X;q)= \sum_{k, s\ge 1}{D^{(\alpha)}_s\bigl(q^k\bigr) \over k \bigl(q^k-1\bigr) q^{k \sigma_{\alpha, s}}} X^{s k} + {D^{(\alpha)}_{-s}\bigl(q^k\bigr) \over k \bigl(q^k-1\bigr) q^{k \sigma_{\alpha, -s}}} X^{- s k} ,
\ee
where $D^{(\alpha)}_{\pm s}(q) \in \IZ[q]$. Equivalently, we can write the coefficient functions from equation~\eqref{phi-struc}~as
 \begin{gather} \label{struc-an}
 a^{(\alpha)}_{\pm n}(q)= \sum_{k|n} {D^{(\alpha)}_{\pm {n \over k}}\bigl(q^k\bigr) \over k\bigl(q^k-1\bigr) q^{k \sigma_{\alpha, \pm {n \over k}}}} , \qquad n \in \IZ_{>0} ,
 \end{gather}
 where the sum runs over all positive integer divisors of $n$.
Since the following discussion holds for all non-abelian choices of the classical branch, let us now drop the explicit dependence on $\alpha$ to simplify the notation and introduce
\be \label{Dtilde1}
\tD_{\pm s}(q)= {D_{\pm s}(q) \over (q-1) q^{\sigma_{\pm s}}} , \qquad s \in \IZ_{>0} .
\ee
Using equations~\eqref{multi-conj} and~\eqref{Dtilde1}, we can write
\begin{equation}
\begin{aligned}[b]
 \exp \bigl( \phi^{\rm WKB}(X;q) \bigr)& = \exp\Bigg( \sum_{k, s\ge 1} {1 \over k} \tD_s\bigl(q^k\bigr) X^{sk} \Bigg) \exp\Bigg( \sum_{k, s\ge 1} {1 \over k} \tD_{-s}\bigl(q^k\bigr) X^{-sk} \Bigg) \\
&= \operatorname{Exp}_{X,q}\Bigg( \sum_{s\ge 1} \tD_s(q) X^s \Bigg) \operatorname{Exp}_{X,q}\Bigg( \sum_{s\ge 1} \tD_{-s}(q) X^{-s} \Bigg) ,\end{aligned}\label{Exp}
\end{equation}
where $\operatorname{Exp}_{X,q}$ denotes the plethystic exponential in the variables $X$, $q$. See, e.g.,~\cite{lm-pv}.
It follows that we can extract the functions \smash{$\tD_{\pm s}(q)$} by applying the plethystic logarithm in the same two variables, which we denote by ${\rm Log}_{X,q}$, to both sides of equation~\eqref{Exp}. Specifically, we find that
\be
\sum_{s\ge 1} \tD_{\pm s}(q) X^{\pm s}={\rm Log}_{X,q} \Bigg( \exp\Bigg( \sum_{n \ge 1} a_{\pm n}(q) X^{\pm n}\Bigg) \Bigg)= \sum_{k, n\ge 1} {\mu(k) \over k} a_{\pm n} \bigl(q^k\bigr) X^{\pm n k} ,
\ee
where $\mu(k)$ is the M\"obius function, which leads to the closed formula
\be \label{Dtilde2}
\tD_{\pm s}(q)= \sum_{k|s} {\mu(k) \over k} a_{\pm {s\over k}} \bigl(q^k\bigr) , \qquad s \in \IZ_{>0} .
\ee
Note that this is the formal inverse of equation~\eqref{struc-an}, which can be equivalently written as
\be \label{struc-an2}
a_{\pm n}(q)= \sum_{k|n} {1 \over k} \tD_{\pm {n\over k}}\bigl(q^k\bigr) , \qquad n \in \IZ_{>0} .
\ee

Let us explain how our conjectures originate from insights from the TS/ST correspondence. See~\cite{mmrev} for a review and further references.
In the context of the TS/ST correspondence, an analogous statement to the AJ conjecture for the CS wave function in equation~\eqref{AJ-eq} arises from the quantization of the mirror curve to a toric Calabi--Yau threefold and the corresponding spectral problem. Its wave function can be computed in the WKB approximation as a formal power series in $\hbar$. It was noted in~\cite{kpamir, mz-wv} that this $\hbar$-expansion could be suitably resummed, producing the same picture we described in our claims above. This structure is similar to the one appearing in perturbative open topological string theory~\cite{lmv,ov}, but it is slightly less constrained. Specifically, one has the structure of~\cite{ov} but not the more detailed substructure found in~\cite{lmv}.
The integrality property in equation~\eqref{multi-conj} characterizes what is called an {\it admissible series} by Kontsevich and Soibelman in~\cite{ks},\footnote{We thank Stavros Garoufalidis for pointing this out to us.} and it is observed in other instances, e.g., as shown in~\cite{as}.\looseness=-1

We can now ask how the resummed perturbative partition functions in equation~\eqref{WKB-conj} relate to the exact partition function $\chi(u;\hbar)$ described in Section~\ref{sec: theory}.
It follows both from physical arguments~\cite{bdp} and from explicit computations~\cite{dimofte-levelk, ggm2,gk-qseries} that the exact wave function can be conjecturally decomposed as
\be \label{exactconj}
\chi(u; \hbar)=\sum_{\alpha} \mC_\alpha \CS^{(\alpha)} (X; q)\CS^{(\alpha)} (X_D;q_D) ,
\ee
where the sum runs over the different non-abelian branches of the classical $A$-polynomial labeled by $\alpha$, $\mC_\alpha$
is an appropriate complex constant, and $\CS^{(\alpha)} (X; q)$ is known as {\it holomorphic block}.
Since the holomorphic block labeled by $\alpha$ must reproduce the perturbative expansion of the wave function around the corresponding classical solution, it can be identified with the resummed WKB series in equation~\eqref{WKB-conj}, that is,
\be \label{block-resummed}
\CS^{(\alpha)} (X; q) = \exp \left(\frac{\ri}{\hbar} s^{(\alpha)}_0(x) + s^{(\alpha)}_1(x) + \phi^{(\alpha), {\rm WKB}}(X;q)\right) .
\ee
Therefore, our conjecture in equation~\eqref{multi-conj} implies that holomorphic blocks in complex CS theory have the stated integrality/multicovering-type property, i.e., they are admissible series in the Kontsevich--Soibelman sense. This could have been anticipated from the relation between holomorphic blocks and open topological string partition functions found in some examples~\cite{bdp}.
Indeed, we will show in Section~\ref{sec: examples} that, starting from the known closed expressions for the holomorphic and antiholomorphic blocks~\cite{bdp,dimofte-levelk,ggm2}, we can successfully verify the integrality structure in equation~\eqref{multi-conj} for all non-abelian choices of the flat connection in the examples of the figure-eight and three-twist knots. Furthermore, we will provide a compact proof of admissibility in the same examples by direct application of a theorem of~\cite{ks} together with well-established $q$-series identities.

Let us comment that the structure in equations~\eqref{exactconj} and~\eqref{block-resummed} is again similar to what has been found in the open TS/ST correspondence~\cite{fgrassi,kpamir, mz-wv, mz-wv2, szabolcs}: the exact wave functions are given by sums of products of the resummed WKB wave functions and their duals.\footnote{In the TS/ST correspondence, one can also define {\it off-shell} wave functions, in which case the dual blocks differ from the original ones.}
 As we mentioned in Section~\ref{sec: intro}, a structure similar to the one in equation~\eqref{phi-struc} has already been conjectured in~\cite{ekholm}, where the resummed WKB expansion is obtained by solving directly the quantized $A$-polynomial equation. The approach of~\cite{ekholm} does not appeal to the decomposition into holomorphic blocks, and our explicit results for the resummation of the WKB expansion in the case of, e.g., the figure-eight knot in Section~\ref{sec: examples} appear to be different from those in~\cite{ekholm}.\footnote{One reason for this difference, which was pointed out to us by Sergei Gukov, is that the results of~\cite{ekholm} involve the super $A$-polynomial, which might lead to a different specialization for $\mathrm{SL}(2,\IC)$.}

\subsection{Cancellation of singularities} \label{sec: cancellation}
Recall that the exact wave function $\chi(u;\hbar)$ is a holomorphic function of $\hbar \in \mathbb{C}'$. In particular, it is well-defined for
\be \label{rathbar}
\hbar = 2 \pi {P \over Q} ,
\ee
where $P$, $Q$ are coprime positive integers. We will refer to values of $\hbar$ of the form in equation~\eqref{rathbar} as {\it rational values}.
On the other hand, due to the conjectured integrality structure in equation~\eqref{multi-conj}, $\phi^{\rm (\alpha), WKB} (X; q)$ is singular precisely for these values of $\hbar$, thus implying that the singularities must disappear in the decomposition of $\chi(u;\hbar)$ in equation~\eqref{exactconj}. More precisely, we will show how these singularities cancel in the sum
\be \label{phi-sum}
\phi^{\rm (\alpha), WKB} (X; q) +\phi^{\rm (\alpha), WKB} (X_D; q_D) ,
 \ee
which implies non-trivial constraints on the coefficients $\hat a_{\pm n}^{(\alpha)}(q)$ in equation~\eqref{ahatf-def}.
This type of branch-by-branch cancellation of singularities found a prototypical example in Faddeev’s quantum dilogarithm~\cite{garou-kas} and played a major role in the understanding of the ABJM matrix model~\cite{hmo2} and the TS/ST correspondence~\cite{ghm,km, wzh}.

Again, we hide the explicit dependence on the classical solution $\alpha$ to simplify the notation. Here and in the rest of this work, we will only reintroduce it when necessary.
We will now study how the cancellation occurs in the exact wave function $\chi(u;\hbar)$ as given in equation~\eqref{exactconj}, and, in Section~\ref{sec: rational1}, we will
compute the finite, well-defined piece that is left from the cancellation and constitutes the sum in equation~\eqref{phi-sum}. Moreover, as we will see in Section~\ref{sec: rational3} along the lines of~\cite{szabolcs}, it is possible to compute this finite part by using only the information contained in the quantum operator $\hat{A}\bigl(\hat{P}, \hat{X},q\bigr)$.
We take
\be
q=\re^{\ri \hbar}=\re^{2 \pi \ri {P \over Q}} , \qquad P,Q \in \IZ_{>0} \ {\rm coprime} ,
\ee
and, substituting into equation~\eqref{phi-struc}, we find that the possible singularities of $\phi^{\rm WKB} (X; q)$ occur when $n=s Q$ and $s \in \IZ_{>0}$. In fact, for these values of $n$, the coefficient functions $a_{\pm n}(q)$ in equation~\eqref{ahatf-def} become
\be \label{ahatsQ}
a_{\pm s Q}(q) = \frac{\hat{a}_{ \pm sQ}( q )}{sQ \bigl(q^{sQ}-1\bigr) q^{ \sigma_{\pm sQ}}} .
\ee
Similarly, considering the dual variable
\be
q_D=\re^{{4 \pi^2 \ri \over \hbar}}=\re^{2 \pi \ri {Q \over P}} ,
\ee
the possible singularities of the dual function $\phi^{\rm WKB} (X_D; q_D)$ occur when $n=s P$ and $s \in \IZ_{>0}$. Indeed, the coefficient functions $a_{\pm n}(q_D)$ in equation~\eqref{ahatf-def} become
\be \label{ahatsP}
a_{\pm s P}( q_D) = \frac{\hat{a}_{\pm sP}( q_D)}{sP \bigl(q_D^{sP}-1\bigr) q_D^{ \sigma_{\pm sP}}} .
\ee
Let us now introduce
\be \label{hbareps}
q_{\e}= \re^{\ri \hbar_{\e}} , \qquad q_{\e, D}= \re^{{ 4 \pi^2 \ri \over \hbar_{\e}}} , \qquad \hbar_{\e}= 2 \pi {P \over Q}+ \epsilon , \qquad 0 < \e \ll 1 ,
\ee
and consider the limit $\e \rightarrow 0$. As detailed in Appendix~\ref{app: eps}, equations~\eqref{ahatsQ} and~\eqref{ahatsP} produce the NLO $\e$-expansions
\begin{subequations}
\begin{gather}
a_{\pm s Q}( q_{\e} ) = \left(- \frac{\ri}{(sQ)^2 \e} -\frac{1}{2sQ} + \frac{1}{(sQ)^2} q \frac{\d }{\d q} \right) \frac{ \hat{a}_{\pm sQ}(q) }{q^{\sigma_{\pm sQ}}} + \CO(\e) , \label{pole} \\
a_{\pm s P}( q_{\e, D} ) = \left( \frac{\ri}{(sQ)^2 \e} +\frac{1}{sP}\left(\frac{\ri}{2 \pi sQ}- {1 \over 2} \right) + \frac{1}{(sP)^2} q_D \frac{\d }{\d q_D} \right) \frac{ \hat{a}_{\pm sP}(q_D)}{q_D^{\sigma_{\pm sP}}} + \CO(\e) , \label{poleD}
\end{gather}
\end{subequations}
respectively. We then require the cancellation of the $\e$-poles in equations~\eqref{pole} and~\eqref{poleD}, which yields the relation
\be
\label{aas}
\hat a_{\pm sQ}\bigl(\re^{2 \pi \ri {P \over Q}} \bigr) \re^{- 2 \pi \ri {P \over Q} \sigma_{\pm sQ}} = \hat a_{\pm sP} \bigl(\re^{2 \pi \ri {Q \over P}} \bigr) \re^{- 2 \pi \ri {Q \over P} \sigma_{\pm sP}} ,
 \ee
for all $s, P, Q \in \IZ_{>0}$ with $P$, $Q$ coprime.
We remark that, for the geometric branch, the formula in equation~\eqref{aas} assumes the simplified form\footnote{A similar cancellation requirement was obtained in~\cite{szabolcs} in the study of the wave function for quantum mirror curves.}
\be
\label{aas-geom}
\hat a^{({\rm geom})}_{sQ}\bigl(\re^{2 \pi \ri {P \over Q}} \bigr) = \hat a^{({\rm geom})}_{sP} \bigl(\re^{2 \pi \ri {Q \over P}} \bigr) ,
 \ee
 where again $s, P, Q \in \IZ_{>0}$ with $P,Q$ coprime.

 Let us now prove that the cancellation formula in equation~\eqref{aas} is a direct consequence of the integrality structure presented in equation~\eqref{multi-conj}. More precisely, equations~\eqref{ahatf-def} and~\eqref{struc-an} imply that
 \be \label{an-dec}
 \hat{a}_n(q) q^{-\sigma_n}= \sum_{k|n} k D_k \bigl( q^{n \over k} \bigr) q^{-{n \over k} \sigma_k} \sum_{j=0}^{k-1} q^{{n \over k}j} ,
 \ee
and, after substituting $n=sQ$ and \smash{$q = \re^{2 \pi \ri {P \over Q}}$} with $s$, $P$, $Q$ as above, we obtain
 \be \label{asQtemp}
 \hat{a}_{sQ}\bigl(\re^{2 \pi \ri {P \over Q}}\bigr) \re^{- 2 \pi \ri {P \over Q} \sigma_{sQ}}= \sum_{k| sQ} k D_k \bigl( \re^{2 \pi \ri \frac{sP}{k}} \bigr) \re^{- 2 \pi \ri \frac{sP}{k} \sigma_{k}} \sum_{j=0}^{k-1} \re^{2 \pi \ri \frac{sP}{k} j} .
 \ee
Since the sum over $j$ on the right-hand side is non-zero if and only if $k | s$, equation~\eqref{asQtemp} gives
\be \label{cs-sequenceP}
 \hat{a}_{sQ}\bigl(\re^{2 \pi \ri {P \over Q}}\bigr) \re^{- 2 \pi \ri {P \over Q} \sigma_{sQ}}= \sum_{k| s} k^2 D_k (1) = c_s \in \IZ , \qquad s \in \IZ_{>0} ,
 \ee
which is independent of the choice of $P,Q \in \IZ_{>0}$ coprime, thus proving the cancellation identity in equation~\eqref{aas}.
We note that the case of $\hat{a}_{-n}(q)$ is handled analogously, yielding the same result. In particular, recalling that $\sigma_{-n}=0$, $n \in \IZ_{>0}$, it defines the sequence of integers
\be \label{cs-sequenceM}
 \hat{a}_{-sQ}\bigl(\re^{2 \pi \ri {P \over Q}}\bigr) = \sum_{k| s} k^2 D_{-k} (1) = c_{-s} \in \IZ , \qquad s \in \IZ_{>0} .
\ee
 Finally, applying the M\"obius inversion formula for arithmetic functions, we deduce that
\be
D_{\pm s}(1)= {1\over s^2} \sum_{k|s} \mu(k) c_{\pm{ s \over k}} , \qquad s \in \IZ_{>0} ,
\ee
where $\mu(k)$ is the M\"obius function.
As we will see, the construction presented here is experimentally verified for the knots ${\bf 4}_1$ and ${\bf 5}_2$ in Section~\ref{sec: examples}.

\subsection{The wave function in the rational case, I} \label{sec: rational1}
We will now apply the results obtained in Section~\ref{sec: cancellation} to derive explicitly the exact wave function~$\chi(u;\hbar)$ in equation~\eqref{exactconj} as a formal power series in $X$, $X^{-1}$ for rational values of $\hbar$.
Let us start from the holomorphic components $\phi^{(\alpha), {\rm WKB}} (X; q)$ and fix $\hbar$ as in equation~\eqref{rathbar}, that~is,%
\be
q=\re^{\ri \hbar}=\re^{2 \pi \ri {P \over Q}} , \qquad P,Q \in \IZ_{>0} \ {\rm coprime} .
\ee
As before, we drop the label $\alpha$ for simplicity and intend the following discussion to hold for each non-abelian choice of the classical branch independently. The singular terms come from the values $n = sQ$ for $s \in \IZ_{>0}$, while choosing $n= sQ+k$ with $1 \le k \le Q-1$ produces only regular terms.
Specifically, the terms involving singularities are obtained by substituting equation~\eqref{pole} into equation~\eqref{phi-struc} and contain the non-singular contributions
\be \label{finitep1}
\sum_{s \ge 1} \left( -{1\over 2} +{1\over sQ} q {\partial \over \partial q} \right) \frac{\hat a_{\pm s Q} (q)}{s Q q^{\sigma_{\pm sQ}}} X^{\pm sQ} .
\ee
The regular terms associated with $n \ne sQ$ are instead
\be \label{finitep2}
\sum_{k=1}^{Q-1} \frac{1}{q^k-1} \sum_{s \ge 0} {\hat a_{\pm (s Q +k)} (q) \over (sQ+k) q^{\sigma_{\pm (sQ+k)}}} X^{\pm (sQ+k)} ,
\ee
where we have used that $q^{sQ+k}= q^k$. Thus, the sum of the quantities in equations~\eqref{finitep1} and~\eqref{finitep2} is the finite part of the function $\phi^{\rm WKB}(X;q)$ evaluated at the rational point in equation~\eqref{rathbar}.
It is convenient to introduce the functions
\begin{subequations} \label{phifunct-parts}
\begin{align}
\varphi^{(+)}_Q(X)&= \sum_{s \ge 1} {\hat a_{s Q} (q) \over sQ q^{\sigma_{sQ}}} X^{sQ} , \\
\varphi^{(-)}_Q(X)&= \sum_{s \ge 1} {\hat a_{-s Q} (q) \over (-sQ) q^{sQ + \sigma_{-sQ}}} X^{-sQ} , \\
\varphi^{(+)}_k(X)&= \sum_{s \ge 0} {\hat a_{s Q +k} (q) \over (sQ+k) q^{\sigma_{sQ+k}}} X^{sQ+k} , \qquad k=1, \dots, Q-1 , \\
\varphi^{(-)}_k(X)&= \sum_{s \ge 0} {\hat a_{-(s Q +k)} (q) \over -(sQ+k) q^{sQ+k + \sigma_{-(sQ+k)}}} X^{-(sQ+k)} , \qquad k=1, \dots, Q-1 ,
\end{align}
\end{subequations}
where we hide the dependence on $q$ to simplify the notation. Moreover, we define
\begin{subequations} \label{phifunct}
\begin{gather}
\varphi_Q(X)= \varphi^{(+)}_Q(X) + \varphi^{(-)}_Q(X) , \\
\varphi_k(X)= \varphi^{(+)}_k(X) - q^k \varphi^{(-)}_k(X) , \qquad k=1, \dots, Q-1 .
\end{gather}
\end{subequations}
Putting equations~\eqref{finitep1} and~\eqref{finitep2} together and suitably identifying the functions just introduced in equation~\eqref{phifunct}, we can write the resummed series in the holomorphic block as
\be \label{phirat}
\phi^{\rm WKB}(X;q)= -{1\over 2} \varphi_Q(X) + q{\partial \over \partial q} \int_0^X \varphi_Q(X') {\rd X '\over X'} +
\sum_{k=1}^{Q-1} { \varphi_k(X) \over q^k-1} .
\ee

Let us now consider the antiholomorphic component $\phi^{\rm WKB} (X_D; q_D)$ and fix $\hbar$ as above. The derivation, in this case, is very similar to the previous one except for one novelty, that is, the appearance of $\log X$ terms, which originate from the $\epsilon$-expansion of the $S$-dual variable
\be
X_D=X^{2 \pi \over \hbar}=X^{Q \over P} .
\ee
Let us add a small positive term $\e$ to the rational value of $\hbar$ as in equation~\eqref{hbareps}.
Then, we consider the limit $\e \rightarrow 0$ and find that
\be \label{XDeps}
X_{\e,D}=X^{2 \pi \over \hbar_{\e}}=X^{{Q \over P} {1 \over 1+ {\epsilon \over 2 \pi} {Q \over P}}}= X^{Q\over P} \left( 1 -{Q^2 \over 2 \pi P^2} \e \log X + \CO\bigl(\e^2\bigr) \right) .
\ee
As described before, the singular terms come from the values $n = sP$ for $s \in \IZ_{>0}$, while choosing~${n= sP+k}$ with $1 \le k \le P-1$ produces only regular terms.
Specifically, the terms involving singularities are obtained by substituting equations~\eqref{poleD} and~\eqref{XDeps} into equation~\eqref{phi-struc} and contain the non-singular contributions
\be \label{finitep1D}
\sum_{s \ge 1} \left( \frac{\ri}{2 \pi sQ}-{1\over 2} +{1\over sP} q_D {\partial \over \partial q_D} \mp \frac{\ri}{2 \pi} \log X \right) \frac{ \hat a_{\pm s P} (q_D) }{s P q_D^{\sigma_{\pm sP}}} X^{\pm sQ} .
\ee
The regular terms associated with $n \ne sP$ are instead
\be \label{finitep2D}
\sum_{k=1}^{P-1} \frac{1}{q_D^k-1} \sum_{s \ge 0} {\hat a_{\pm (s P +k)} (q_D) \over (sP+k) q_D^{\sigma_{\pm (sP+k)}}} X^{\pm (sP+k){Q \over P}} ,
\ee
where we have used that $q_D=\re^{2 \pi {Q \over P}}$ and thus $q_D^{sP+k}= q_D^k$.
Therefore, the sum of the quantities in equations~\eqref{finitep1D} and~\eqref{finitep2D} is the finite part of the function $\phi^{\rm WKB}(X_D;q_D)$ evaluated at the rational point in equation~\eqref{rathbar}.
As we have done for the holomorphic part, we introduce the functions
\begin{subequations} \label{phitildefunct-parts}
\begin{align}
\widetilde \varphi^{(+)}_P(X_D)&= \sum_{s \ge 1} {\hat a_{s P} (q_D) \over sP q_D^{\sigma_{sP}}} X_D^{sP} ,\\
\widetilde \varphi^{(-)}_P(X_D)&= \sum_{s \ge 1} {\hat a_{-s P} (q_D) \over -(sP) q_D^{sP+\sigma_{-sP}}} X_D^{-sP} ,\\
\widetilde \varphi^{(+)}_k(X_D)&= \sum_{s \ge 0} {\hat a_{s P +k} (q_D) \over (sP+k) q_D^{\sigma_{sP+k}}} X_D^{sP+k} , \qquad k=1, \dots, P-1 , \\
\widetilde \varphi^{(-)}_k(X_D)&= \sum_{s \ge 0} {\hat a_{-(s P +k)} (q_D) \over -(sP+k) q_D^{sP+k+\sigma_{-(sP+k)}}} X_D^{-(sP+k)} , \qquad k=1, \dots, P-1 ,
\end{align}
\end{subequations}
where we hide the dependence on $q_D$ to simplify the notation. Moreover, we define
\begin{subequations} \label{phitildefunct}
\begin{gather}
\widetilde \varphi_P(X_D)= \widetilde \varphi^{(+)}_P(X_D) + \widetilde \varphi^{(-)}_P(X_D) , \\
\widetilde \varphi_k(X_D)= \widetilde \varphi^{(+)}_k(X_D) - q_D^k \widetilde \varphi^{(-)}_k(X_D) , \qquad k=1, \dots, Q-1 .
\end{gather}
\end{subequations}
Putting equations~\eqref{finitep1D} and~\eqref{finitep2D} together and suitably identifying the functions just introduced in equation~\eqref{phitildefunct}, we can now write the resummed series in the antiholomorphic block~as%
\be \label{phiratD}
\begin{aligned}[b]
\phi^{\rm WKB}(X_D;q_D)={}& -\left( {1\over 2} + {\ri \over 2 \pi} \log X \right) \widetilde \varphi_P(X_D) \\
&+ \left({\ri P \over 2 \pi Q} + q_D{\partial \over \partial q_D} \right) \int_0^{X_D} \widetilde \varphi_P(X_D') {\rd X_D' \over X_D'} + \sum_{k=1}^{P-1} { \widetilde \varphi_k(X_D)\over q_D^k-1} .
\end{aligned}
\ee

Finally, we observe that
\be \label{rel-phi}
\widetilde \varphi^{(+)}_P(X_D)={Q \over P} \varphi^{(+)}_Q(X) , \qquad \widetilde \varphi^{(-)}_P(X_D)={Q \over P} q^{s Q \left( 1- \frac{Q}{P} \right)} \varphi^{(-)}_Q(X) ,
\ee
as a direct consequence of the cancellation symmetry in equation~\eqref{aas}.
Therefore, after summing equations~\eqref{phirat} and~\eqref{phiratD} and using the relations in equation~\eqref{rel-phi}, we obtain the expression
\begin{gather}
\phi^{\rm WKB}(X;q)+ \phi^{\rm WKB}(X_D;q_D)\nonumber\\
\qquad= \left( -{1 \over 2} - {Q \over 2 P} - {\ri Q \over 2 \pi P} \log X \right) \varphi_Q(X) + \left( {\ri Q \over 2 \pi P} + 2 q {\partial \over \partial q} \right)\int_0^X \varphi_Q(X') {\rd X '\over X'} \nonumber\\
\phantom{\qquad= }{}+ s Q \left( 1- \frac{Q}{P} \right) \int_0^X \varphi^{(-)}_Q(X') {\rd X '\over X'} + \sum_{k=1}^{Q-1} { \varphi_k(X)\over q^k-1} + \sum_{k=1}^{P-1} { \widetilde \varphi_k\bigl(X^{Q \over P}\bigr)\over q^{k{Q^2 \over P^2}}-1} .\label{phirat-final}
\end{gather}
Besides, the power series newly defined in equations~\eqref{phifunct} and~\eqref{phitildefunct}, which dictate the formula for the exact wave function at rational points by means of equation~\eqref{phirat-final}, are explicitly and uniquely determined by the function
\be \label{upsilon}
\Upsilon (X)=\Upsilon^{(+)} (X)+\Upsilon^{(-)} (X)=\sum_{n \ge 1} {\hat{a}_n (q) \over n q^{\sigma_n}} X^n + \sum_{n \ge 1} {\hat{a}_{-n} (q) \over (-n) q^{n+\sigma_{-n}}} X^{-n} ,
\ee
where, again, we hide the dependence on $q$ to simplify the notation. Indeed, we can write
\begin{subequations}
\begin{gather}
\varphi_Q(X)={1\over Q} \sum_{m=0}^{Q-1} \Upsilon (q^m X) , \label{varphiQ-U} \\
\varphi^{(\pm)}_k(X)={1\over Q} \sum_{m=0}^{Q-1} q^{\mp m k} \Upsilon^{(\pm)} (q^m X,q) , \qquad k=1, \dots, Q-1 ,
\end{gather}
\end{subequations}
and similarly
\begin{subequations}
\begin{gather}
\widetilde \varphi_P(X_D)={1\over P} \sum_{m=0}^{P-1} \Upsilon (q_D^m X_D) , \\
\widetilde \varphi^{(\pm)}_k(X_D)={1\over P} \sum_{m=0}^{P-1} q_D^{\mp m k} \Upsilon^{(\pm)} (q_D^m X_D) , \qquad k=1, \dots, P-1 .
\end{gather}
\end{subequations}
Let us stress that the derivation we have presented here implies a fixed choice of the classical branch. Indeed, the functions $\Upsilon (X)$, $\varphi_Q(X)$, $\varphi_k(X)$, $\widetilde \varphi_P(X_D)$, $\widetilde \varphi_k(X_D)$ are implicitly dependent on the label $\alpha$.
We note that, applying equations~\eqref{cs-sequenceP} and~\eqref{cs-sequenceM}, we can simply read off the integer coefficients $c_{\pm s}$, $s \in \IZ_{>0}$, from the function $\varphi_Q(X)$. Namely, we have that
\be \label{varphiQ-cs}
\varphi_Q(X)=\sum_{s \ge 1} {c_s \over sQ} X^{sQ} - {c_{-s} \over sQ} X^{-sQ} ,
\ee
after substituting $q^{sQ}=1$.
In Appendix~\ref{app: rational2}, we describe an alternative way of computing the integer sequences $\{ c_{\pm s} \}$, $s \in \IZ_{>0}$ using equation~\eqref{phirat-final}.
We will show in Section~\ref{sec: examples} that, starting from the known closed expression for the holomorphic and antiholomorphic blocks, we can successfully compute the integer coefficients for both series in $X$, $X^{-1}$ and all choices of the flat connection $\alpha$ in the examples of the figure-eight and three-twist knots.

\subsection{The wave function in the rational case, II} \label{sec: rational3}
In Section~\ref{sec: rational1}, we showed that the exact wave function $\chi(u; \hbar)$ in equation~\eqref{exactconj} at rational values of $\hbar$ is determined by the functions $\Upsilon^{(\alpha)} (X)$, defined in equation~\eqref{upsilon}, and their derivatives with respect to $q$, up to the constants $\mC_\alpha$ and the exponential prefactors containing \smash{$s^{(\alpha)}_{0,1}(x)$}. We will now prove that the functions \smash{$\Upsilon^{(\alpha)} (X)$} can be obtained by solving directly the quantum $A$-polynomial in closed form. We will focus on $q$-difference equations of order two and three. These are the relevant cases for the examples studied in Section~\ref{sec: examples}. The matrix formalism presented below was used in~\cite{szabolcs} for a second-order $q$-difference equation in the context of quantum mirror curves.

\subsubsection[Solving a second-order q-difference equation]{Solving a second-order $\boldsymbol{ q}$-difference equation} \label{sec: 2nd-order}
Recall that, for each choice of $\alpha$, the resummed perturbative wave function $\chi^{(\alpha), {\rm WKB}}(u; \hbar)$ is given in equation~\eqref{WKB-conj}. Equivalently, we write
\be \label{chi-matrix-2}
\chi^{(\alpha), {\rm WKB}}(u; \hbar) = \re^{\frac{\ri}{\hbar} s^{(\alpha)}_0(x) + s^{(\alpha)}_1(x)} \Psi^{(\alpha), {\rm WKB}}(X;q) ,
\ee
where we have introduced the notation
\be \label{Psi}
\Psi^{(\alpha), {\rm WKB}}(X;q) = \exp\bigl( \phi^{(\alpha), {\rm WKB}}(X;q)\bigr) ,
\ee
which will be useful in the following discussion. Let us assume that the corresponding quantum $A$-polynomial equation, that is,
\be \label{Aeq-matrix}
\hat{A}\bigl(\hat{P}, \hat{X}, q\bigr) \chi^{(\alpha), {\rm WKB}}(u; \hbar) = 0 ,
\ee
is of second order. This is the case of the $\FE$-knot, as shown in equation~\eqref{rec41x}. After appropriately accounting for the correction due to the exponential prefactor containing the functions~\smash{$s^{(\alpha)}_{0,1}(x)$} in equation~\eqref{chi-matrix-2}, the $q$-difference equation above becomes
\be \label{Psi-eq}
\Psi^{(\alpha), {\rm WKB}}(q X;q)+A(X)\Psi^{(\alpha), {\rm WKB}}(X;q)+B(X)\Psi^{(\alpha), {\rm WKB}}\bigl(q^{-1} X;q\bigr)=0 ,
\ee
where $A(X)$, $B(X)$ are functions of $X$ and $q$.
Let us observe that, due to the conjectural structure in equations~\eqref{phi-struc} and~\eqref{ahatf-def}, we have that
\be \label{uU-def}
 \upsilon^{(\alpha)} (X) = {\Psi^{(\alpha), {\rm WKB}} (q X;q) \over \Psi^{(\alpha), {\rm WKB}}(X;q)}=\exp\bigl( \Upsilon^{(\alpha)} (X) \bigr) ,
\ee
and we can now write equation~\eqref{Psi-eq} in terms of the above function $\upsilon^{(\alpha)} (X)$ as
\be \label{VVab}
\upsilon (X)\upsilon\bigl(q^{-1} X\bigr)+A(X) \upsilon\bigl(q^{-1} X\bigr)+B(X)=0 ,
\ee
where we have dropped the label $\alpha$ from the notation for simplicity.
Let us further assume that~$\hbar$ is fixed as in equation~\eqref{rathbar}, that is,
\be \label{q-rat}
q=\re^{\ri \hbar}=\re^{2 \pi \ri {P \over Q}} , \qquad P,Q \in \IZ_{>0} \ {\rm coprime} ,
\ee
and introduce the sequence of functions $\{\upsilon_k \}$, $k \in \IZ_{\ge 0}$, of $X$ and $q$ defined by
\be
\upsilon_k = \upsilon \bigl(q^{-k} X\bigr) .
\ee
They satisfy the system of equations
\be \label{vkvk1}
\upsilon_k \upsilon_{k+1}+A\bigl(q^{-k} X\bigr) \upsilon_{k+1}+B\bigl(q^{-k} X\bigr)=0 , \qquad 0 \le k \le Q-1 ,
\ee
as a consequence of equation~\eqref{VVab}. We solve this system recursively, as follows. We define the sequences $\big\{ a^{(k)}(X) \big\}$ and $\big\{ b^{(k)}(X) \big\}$, $k \in \IZ_{\ge 0}$, via the relations
\begin{subequations} \label{ABvprods}
\begin{align}
	 &a^{(1)}(X) =A(X),\\
	 & b^{(1)}(X) =B(X), \\
	 &\upsilon_0 \upsilon_1 \cdots \upsilon_{k}+a^{(k)}(X) \upsilon_{k}+b^{(k)}(X) =0 , \qquad k > 1 . \label{ABvprods-last}
\end{align}
\end{subequations}
If we multiply equation~\eqref{ABvprods-last} by $\upsilon_{k+1}$ and apply equation~\eqref{vkvk1}, we obtain a recursion relation which can be written in
matrix form as
\be
	\begin{pmatrix}
		a^{(k+1)}(X) \\
		b^{(k+1)}(X)
	\end{pmatrix}
	=
	\begin{pmatrix}
		-A\bigl(q^{-k} X\bigr) & 1 \\
		-B\bigl(q^{-k} X\bigr) & 0
	\end{pmatrix}
	\begin{pmatrix}
		a^{(k)}(X) \\
		b^{(k)}(X)
	\end{pmatrix} ,
\ee
and it can be solved as
\be
	\label{relQm1}
	\begin{pmatrix}
		a^{(Q)}(X) \\
		b^{(Q)}(X)
	\end{pmatrix}
	=
	{\mathcal M}_Q(X)
	\begin{pmatrix}
		-1 \\
		0
	\end{pmatrix} .
\ee
Here, ${\mathcal M}_Q(X)$ is the matrix
\be \label{matrix2}
\CM_Q(X)	
	=
	\prod_{k=1}^{Q}
	\begin{pmatrix}
		-A\bigl(q^k X\bigr) & 1 \\
		-B\bigl(q^k X\bigr) & 0
	\end{pmatrix} =
	\begin{pmatrix}
		\CM_{11}(X) & \CM_{12}(X) \\
		\CM_{21}(X) & \CM_{22}(X)
	\end{pmatrix} ,
\ee
where the product is ordered from left to right as $k$ increases.
Recalling that $q^Q=1$, we obtain
\begin{align}
	{\mathcal M}_Q\bigl(q^{-1} X\bigr) &=
	\begin{pmatrix}
		-A(X) & 1 \\
		-B(X) & 0
	\end{pmatrix}
	{\mathcal M}_Q(X)
	\begin{pmatrix}
		-A( X) & 1 \\
		-B(X) & 0
	\end{pmatrix}^{-1} \nonumber\\
	&=
	\begin{pmatrix}
		\CM_{22} -A \CM_{12} & \frac{A}{B} (\CM_{11}+A \CM_{12} ) - \frac{1}{B} ( \CM_{21}+A \CM_{22} ) \\
		-B \CM_{12} & \CM_{11}+A \CM_{12}
	\end{pmatrix} ,\label{MqM}
\end{align}
where we have removed the explicit dependence on $X$ for compactness.
Following equations \eqref{relQm1} and~\eqref{MqM}, we find that\footnote{There is a sign misprint in the corresponding equa\-tion in~\cite{szabolcs}.}
\be \label{abM2}
	\begin{pmatrix}
		a^{(Q)}\bigl(q^{-1}X\bigr) \\
		b^{(Q)}\bigl(q^{-1}X\bigr)
	\end{pmatrix}
	=\begin{pmatrix}
		{-\mathcal M}_{22}(X)+A(X){\mathcal M}_{12}(X) \\
		B(X){\mathcal M}_{12}(X)
	\end{pmatrix} .
\ee
Let us now define the function
\be
	\Pi \upsilon = \upsilon_0 \upsilon_1 \cdots \upsilon_{Q-1} ,
\ee
which is invariant under $q$-shifts. We consider equation~\eqref{ABvprods-last} for $k=Q$ and, shifting $X$ into~$q^{-1}X$, we produce the system
	\begin{align}
	&\upsilon_0\bigl(\Pi \upsilon+ a^{(Q)}(X)\bigr)+b^{(Q)}(X)=0 , \qquad
	 \upsilon_1\bigl(\Pi \upsilon+a^{(Q)}\bigl(q^{-1}X\bigr)\bigr)+b^{(Q)}\bigl(q^{-1}X\bigr)=0 .
	\end{align}
By substituting the expressions for $a^{(Q)}(X)$, $a^{(Q)}\bigl(q^{-1}X\bigr)$ and $b^{(Q)}(X)$, $b^{(Q)}\bigl(q^{-1}X\bigr)$ in terms of the entries of the original matrix $\mathcal M_Q(X)$ in equations~\eqref{relQm1} and~\eqref{abM2} and then using equation~\eqref{vkvk1} with $k=0$, we can rewrite the system above as
\be
	\upsilon_0(\Pi \upsilon- {\mathcal M}_{11}(X))- {\mathcal M}_{21}(X)=0 , \qquad
	\Pi \upsilon - {\mathcal M}_{22}(X)-\upsilon_0 {\mathcal M}_{12}(X)=0 ,\label{v0system}
\ee
or, equivalently, in the matrix form
\be
	\begin{pmatrix}
		{\mathcal M}_{11}(X)-\Pi \upsilon & {\mathcal M}_{21}(X) \\
		{\mathcal M}_{12}(X) & {\mathcal M}_{22}(X)-\Pi \upsilon
	\end{pmatrix}
	\begin{pmatrix}
		\upsilon_0 \\
		1
	\end{pmatrix}
	=
	\begin{pmatrix}
		0 \\
		0
	\end{pmatrix} .
\ee
It follows straightforwardly that $\Pi \upsilon$ is an eigenvalue of the transpose matrix $\mathcal M_Q(X)^{\rm T}$ with eigenvector $(\u_0, 1)^{\rm T}$. Specifically, we have that
\be \label{Piu-2}
\Pi \upsilon = \frac{ \operatorname{tr}{\mathcal M_Q}(X) \pm \sqrt{\Delta_Q(X)} }{2} ,
\ee
where we have introduced
\be
\Delta_Q(X) = (\operatorname{tr}{\mathcal M_Q}(X))^2 - 4 \det {\mathcal M_Q}(X) ,
\ee
while $\det {\mathcal M_Q}(X)$ and $\operatorname{tr}{\mathcal M_Q}(X) $ denote the trace and determinant of the matrix, respectively.
Note that, following equation~\eqref{matrix2}, both $\det {\mathcal M_Q}(X)$ and $\operatorname{tr}{\mathcal M_Q}(X) $ are invariant under $q$-shifts and so depend on $X$ through $X^Q$.
The function $\upsilon_0$ can be found, for example, using the second line of equation~\eqref{v0system}, yielding
\be \label{uo-2}
\upsilon_0 = \frac{ \Pi v - {\mathcal M}_{22}(X) }{{\mathcal M}_{12}(X)} = \frac{ {\mathcal M}_{11}(X) - {\mathcal M}_{22}(X) \pm \sqrt{\Delta_Q(X)} }{2 {\mathcal M}_{12}(X)} ,
\ee
while the functions $\upsilon_k$, $k \in \IZ_{>0}$, are obtained by successively $q$-shifting $\upsilon_0$. This gives the solution to equation~\eqref{VVab} for all rational values of $\hbar$.
However, we stress that, for such a~solution to be consistent with the original choice of a classical branch labeled by $\alpha$, we must impose that, taking $P=Q=1$, we find
\be \label{branch}
\u_0= \Pi \u = P^{(\alpha)} ,
\ee
where $P^{(\alpha)}$ is the selected non-abelian solution of the classical $A$-polynomial at fixed $X$ after taking into account the corrections due to the exponential prefactor in equation~\eqref{chi-matrix-2}. In this way, we remove the sign ambiguity in equation~\eqref{Piu-2} and correspondingly in equation~\eqref{uo-2}.
We conclude by observing that equation~\eqref{uU-def} implies
\begin{subequations} \label{2nd-end}
\begin{align}
	\Upsilon (X) &= \log \upsilon_0 = \log \left ( \frac{ {\mathcal M}_{11}(X)- {\mathcal M}_{22}(X) \pm \sqrt{\Delta_Q(X)} }{2 {\mathcal M}_{12}(X)} \right ) , \\
	\varphi_Q(X) &= \frac{1}{Q} \log \Pi\upsilon = \frac{1}{Q} \log \left ( \frac{ \operatorname{tr}{\mathcal M}_Q(X) \pm \sqrt{\Delta_Q(X)} }{2} \right ) ,
\end{align}
\end{subequations}
where we have applied the relation
\be \label{varphi-Unew}
\Pi \u = \exp \left( \sum_{m=0}^{Q-1} \Upsilon(q^m X) \right) = \exp (Q \varphi_Q (X) ) ,
\ee
which is a direct consequence of equation~\eqref{varphiQ-U}. Again, the sign ambiguity in equation~\eqref{2nd-end} is resolved by the requirement in equation~\eqref{branch}.

\subsubsection[Solving a third-order q-difference equation]{Solving a third-order $\boldsymbol{ q}$-difference equation} \label{sec: 3rd-order}
We will show how the matrix formalism of Section~\ref{sec: 2nd-order} can be applied to solve a third-order $q$-difference equation in closed form.
In particular, let us return to the quantum $A$-polynomial equation satisfied by the resummed perturbative wave function $\chi^{(\alpha), {\rm WKB}}(u; \hbar)$ in equation~\eqref{Aeq-matrix} and assume it is of order three. This is the case of the $\TT$-knot, as shown in equation~\eqref{rec52x}.
As before, after appropriately accounting for the correction due to the exponential prefactor containing the functions \smash{$s^{(\alpha)}_{0,1}(x)$} in equation~\eqref{chi-matrix-2}, the $q$-difference equation takes the form
\begin{gather}
\Psi^{(\alpha), {\rm WKB}}(q X;q)+A(X)\Psi^{(\alpha), {\rm WKB}}(X;q)+B(X)\Psi^{(\alpha), {\rm WKB}}\bigl(q^{-1} X;q\bigr) \nonumber\\
\qquad{}+C(X)\Psi^{(\alpha), {\rm WKB}}\bigl(q^{-2} X;q\bigr)=0 ,\label{Psi-eq3}
\end{gather}
where $A(X)$, $B(X)$, $C(X)$ are functions of $X$ and $q$. We can now introduce the function $\u^{(\alpha)}(X)$ as in equation~\eqref{uU-def} and write equation~\eqref{Psi-eq3} equivalently as
\be \label{VVab3}
\upsilon (X) \upsilon\bigl(q^{-1} X\bigr) \upsilon\bigl(q^{-2} X\bigr) + A(X) \upsilon\bigl(q^{-1} X\bigr)\upsilon\bigl(q^{-2} X\bigr) + B(X) \upsilon\bigl(q^{-2} X\bigr) + C(X) = 0 ,
\ee
where we have dropped the label $\alpha$ from the notation for simplicity.
Let us further assume that~$q$ is fixed as in equation~\eqref{q-rat} and define the sequence of functions $\{\upsilon_k \}$, $k \in \IZ_{\ge 0}$, of $X$ and~$q$~by
\be
\upsilon_k = \upsilon \bigl(q^{-k} X\bigr) .
\ee
It follows from shifting $X$ into $q^{-k}X$ in equation~\eqref{VVab3} that they satisfy the system of equations
\begin{gather}
\u_k \u_{k+1} \u_{k+2} + A\bigl(q^{-k} X\bigr) \u_{k+1} \u_{k+2} + B\bigl(q^{-k} X\bigr) \u_{k+2} + C\bigl(q^{-k} X\bigr) = 0 , \nonumber\\
 0 \le k \le Q-1 .\label{vkvk13}
\end{gather}
In order to solve the above system recursively, let us define the sequences $\big\{ a^{(k)}(X)\big\}$, $\big\{b^{(k)}(X)\big\}$, and $\big\{c^{(k)}(X)\big\}$, $k \in \IZ_{\ge 0}$, via the relations
\begin{subequations} \label{ABvprods3}
\begin{align}
&a^{(1)}(X) = A(X) , \\
&b^{(1)}(X) = B(X) , \\
&c^{(1)}(X) = C(X) , \\
&\u_0 \u_1 \cdots \u_{k+1} + a^{(k)}(X) \u_k \u_{k+1} + b^{(k)}(X) \u_{k+1} + c^{(k)}(X)=0 , \qquad k>1 . \label{ABvprods3-last}
\end{align}
\end{subequations}
If we multiply equation~\eqref{ABvprods3-last} by $\u_{k+2}$ and apply equation~\eqref{vkvk13}, we obtain a recursion relation which can be written in matrix form as
\be
	\begin{pmatrix}
		a^{(k+1)}(X) \\
		b^{(k+1)}(X) \\
		c^{(k+1)}(X)
	\end{pmatrix}
	=
	\begin{pmatrix}
		-A\bigl(q^{-k} X\bigr) & 1 & 0 \\
		-B\bigl(q^{-k} X\bigr) & 0 & 1 \\
		-C\bigl(q^{-k} X\bigr) & 0 & 0
	\end{pmatrix}
	\begin{pmatrix}
		a^{(k)}(X) \\
		b^{(k)}(X) \\
		c^{(k)}(X)
	\end{pmatrix} ,
\ee
and it can be solved as
\be
	\label{relQm13}
	\begin{pmatrix}
		a^{(Q)}(X) \\
		b^{(Q)}(X) \\
		c^{(Q)}(X)
	\end{pmatrix}
	=
	{\mathcal M}_Q(X)
	\begin{pmatrix}
		-1 \\
		0 \\
		0
	\end{pmatrix} .
\ee
Here, ${\mathcal M}_Q(X)$ is the matrix
\be \label{matrix3}
\CM_Q(X)	
	=
	\prod_{k=1}^{Q}
	\begin{pmatrix}
		-A\bigl(q^k X\bigr) & 1 & 0 \\
		-B\bigl(q^k X\bigr) & 0 & 1 \\
		-C\bigl(q^k X\bigr) & 0 & 0
	\end{pmatrix}
	=
	\begin{pmatrix}
		\CM_{11}(X) & \CM_{12}(X) & \CM_{13}(X) \\
		\CM_{21}(X) & \CM_{22}(X) & \CM_{23}(X) \\
		\CM_{31}(X) & \CM_{32}(X) & \CM_{33}(X)
	\end{pmatrix} ,
\ee
where the product is ordered from left to right as $k$ increases. As before, let us introduce
\be
	\Pi \upsilon = \upsilon_0 \upsilon_1 \cdots \upsilon_{Q-1} ,
\ee
which is invariant under $q$-shifts. Moreover, as it will be useful in the following, we compute the $q$-shifted matrix
\be
\CM_Q\bigl(q^{-1} X\bigr) =
\begin{pmatrix}
 -A(X) & 1 & 0 \\
 -B(X) & 0 & 1 \\
 -C(X) & 0 & 0
\end{pmatrix}
\CM_Q(X)
\begin{pmatrix}
 -A(X) & 1 & 0 \\
 -B(X) & 0 & 1 \\
 -C(X) & 0 & 0
\end{pmatrix}^{-1} ,
\ee
which can be written as
\begin{gather} \label{dict}
 \left(\!\begin{matrix}
 -A \CM_{12} \!+\! \CM_{22}\! &\! -A \CM_{13}\! +\! \CM_{23} & \!\frac{A}{C} (\CM_{11}\!+\! A \CM_{12}\!+\! B \CM_{13}) \!-\! \frac{1}{C} (\CM_{21}\!+\! A \CM_{22}\!+\! B \CM_{23}) \vspace{1mm}\\
 -B \CM_{12} \!+\! \CM_{32}\! &\! -B \CM_{13}\! +\! \CM_{33} & \!\frac{B}{C} (\CM_{11}\!+\! A \CM_{12}\!+\! B \CM_{13}) \!-\! \frac{1}{C} (\CM_{21}\!+\! A \CM_{22}\!+\! B \CM_{23}) \vspace{1mm}\\
 -C \CM_{12} & -C \CM_{13} & (\CM_{11}\!+\! A \CM_{12}\!+\! B \CM_{13})
\end{matrix} \!\right)\!,
\end{gather}
where we have removed the explicit dependence on $X$ for compactness.
Let us now take equation~\eqref{ABvprods3-last} for $k=Q$, which gives
\be \label{start3}
\u_0 \u_1 \bigl(\Pi \u + a^{(Q)}(X)\bigr) + b^{(Q)}(X) \u_1 + c^{(Q)}(X) = 0 .
\ee
Equation~\eqref{relQm13} allows us to express the functions $a^{(Q)}(X)$, $b^{(Q)}(X)$, and $c^{(Q)}(X)$ in terms of the entries of the matrix $\mathcal M_Q(X)$, so that equation~\eqref{start3} becomes
\be \label{case1}
\u_0 \u_1 (\Pi \u - \CM_{11}(X)) - \CM_{21}(X) \u_1 - \CM_{31}(X) = 0 .
\ee
Shifting $X$ into $q^{-1}X$, using equation~\eqref{vkvk13} with $k=0$, and applying the dictionary encoded in the matrix in equation~\eqref{dict}, we find the second equation
\be \label{case2}
\u_1 (\Pi \u - \CM_{22}(X)) - \CM_{32}(X) - \CM_{12}(X) \u_0 \u_1 = 0 .
\ee
Subsequently shifting $X$ into $q^{-1}X$ in the equation above, again using equations~\eqref{vkvk13} and \eqref{dict}, we obtain the third equation
\be \label{case3}
\Pi \u - \CM_{33}(X) - \CM_{23}(X) \u_1 - \CM_{13}(X) \u_0 \u_1 = 0 .
\ee
Equations~\eqref{case1},~\eqref{case2}, and~\eqref{case3} assemble into a cubic system which can be written in matrix form as
\be
\begin{pmatrix}
\CM_{11}(X) - \Pi \u & \CM_{21}(X) & \CM_{31}(X) \\
\CM_{12}(X) & \CM_{22}(X) - \Pi \u & \CM_{32}(X) \\
\CM_{13}(X) & \CM_{23}(X) & \CM_{33}(X) - \Pi \u
\end{pmatrix}
\begin{pmatrix}
\u_0 \u_1 \\
\u_1 \\
1
\end{pmatrix} =
\begin{pmatrix}
0 \\
0 \\
0
\end{pmatrix} .
\ee
We conclude that $\Pi \u$ is an eigenvalue of the transpose matrix $\CM_Q(X)^{\rm T}$ with eigenvector $(\u_0 \u_1, \u_1, 1)^{\rm T}$. The characteristic polynomial is
\be
p(z) = \det \bigl(z \mathbb{I}_3 - \CM_Q^{\rm T}\bigr) = z^3 + d_2 z^2 + d_1 z + d_0 ,
\ee
where $\mathbb{I}_3$ is the $3\times3$ identity matrix and
\begin{subequations}
\begin{align}
d_2=& - \CM_{11} - \CM_{22} - \CM_{33} , \\
d_1=& - \CM_{23} \CM_{32} - \CM_{12} \CM_{21} - \CM_{13} \CM_{31} + \CM_{11} \CM_{22} + \CM_{11} \CM_{33} + \CM_{22} \CM_{33} , \\
d_0=& - \CM_{11} \CM_{22} \CM_{33} + \CM_{11} \CM_{23} \CM_{32} + \CM_{22} \CM_{31} \CM_{13} + \CM_{33} \CM_{21} \CM_{12} \nonumber\\
& - \CM_{21} \CM_{13} \CM_{32} - \CM_{31} \CM_{12} \CM_{23} .
\end{align}
\end{subequations}
The eigenvalues of the matrix $\CM_Q(X)^{\rm T}$ are the roots of $p(z)$, that is,
\be \label{Pi3}
\Pi \u = -\frac{1}{3} \left( d_2 + \Theta_2 + \frac{\Theta_0}{\Theta_2} \right) ,
\ee
where we have
\begin{subequations}
\begin{align}
\Theta_0 &= d_2^2 - 3 d_1 , \\
\Theta_1 &= 2 d_2^3 - 9 d_2 d_1 + 27 d_0 , \\
\Theta_2 &= { \re^{m {\pi \ri \over 3}} \over 2} \sqrt[3]{\Theta_1+\sqrt{\Theta_1^4 - 4 \Theta_0^3}} , \qquad m=0, 1, 2 . \label{theta2-m}
\end{align}
\end{subequations}
The function $\u_0$ is obtained, for example, by solving the system of equations~\eqref{case2} and~\eqref{case3}, yielding
\be \label{u03}
\u_0 = \frac{\CM_{22} \CM_{33}-\CM_{22} \Pi \u -\CM_{23} \CM_{32}-\CM_{33} \Pi \u+ (\Pi \u)^2}{-\CM_{12} \CM_{33}+\CM_{12} \Pi \u +\CM_{13} \CM_{32}} ,
\ee
while the functions $\upsilon_k$, $k \in \IZ_{>0}$, are computed by successively $q$-shifting $\upsilon_0$.
This gives the solution to equation~\eqref{VVab3} for all rational values of $\hbar$.
As we have done in Section~\ref{sec: 2nd-order}, we can now fix the ambiguity in equation~\eqref{Pi3} and correspondingly in equation~\eqref{u03}, that is, the value of $m=0,1,2$ in the definition of $\Theta_2$, by imposing the consistency of the solution with the original choice of a non-abelian classical branch labeled by $\alpha$. Specifically, we require that equation~\eqref{branch} holds in the case of $P=Q=1$ after accounting for the exponential prefactor in equation~\eqref{chi-matrix-2}.
Again, we conclude by observing that equations~\eqref{uU-def} and~\eqref{varphi-Unew} imply
\be \label{dict-end}
	\Upsilon (X) = \log \upsilon_0 , \qquad \varphi_Q(X) = \frac{1}{Q} \log \Pi\upsilon .
\ee
Note that the procedure described here for a third-order $q$-difference equation can be generalized to higher orders.

\section{Examples} \label{sec: examples}
In this section, we will illustrate how the conjectures and computational methods of Section~\ref{sec: main} apply to the two simplest hyperbolic knots, i.e., the figure-eight ($\FE$) and three-twist ($\TT$) knots.
We will first verify the integrality structure of equation~\eqref{multi-conj} starting from the known decomposition into holomorphic and antiholomorphic blocks.
We will then determine the wave function at rational values of $\hbar$ following the matrix formalism of Section~\ref{sec: rational3} and cross-check the agreement of our results.
Finally, we will include a short proof of admissibility for the examples at hand based on a theorem of Kontsevich and Soibelman~\cite{ks}.

\subsection{The figure-eight knot} \label{sec: 41}
The state integral of the figure-eight knot $\chi_{{\bf{4}}_1}(u;\hbar)$ can be expressed as~\cite{bdp,dimofte-levelk,ggm2}
 \begin{align}
 \chi_{{\bf{4}}_1}(u;\hbar)={}&\re^{-\frac{\pi\ri}{12}-\frac{2\pi\ri}{3}c_\mb^2} \bigl(
\re^{\pi\ri(-2u^2-4c_\mb u)}
 \CJ\bigl(\re^{4\pi\mb u},\re^{2\pi\mb u};q\bigr)
 \CJ\bigl(\re^{4\pi\mb^{-1}u},\re^{2\pi\mb^{-1} u};\tq^{-1}\bigr)\nonumber\\
& +\re^{\pi\ri(u^2-2c_\mb u)}
 \CJ\bigl(\re^{2\pi\mb u},\re^{-2\pi\mb u};q\bigr)
 \CJ\bigl(\re^{2\pi\mb^{-1}u},\re^{-2\pi\mb^{-1} u};\tq^{-1}\bigr)
 \bigr) ,\label{chi41}
\end{align}
where $c_{\mb}= \ri \bigl(\mb + \mb^{-1})/2$ and the $q$-special function $\CJ(x,y;q)$ is defined by\footnote{$\CJ(x,y;q)$ is closely related to the Hahn--Exton $q$-Bessel function~\cite{swarttouw2,swarttouw}.}
\begin{equation} \label{J}
\CJ(x,y;q) = (y q;q)_\infty \sum_{n=0}^\infty (-1)^n \frac{q^{n(n+1)/2} x^n}{(q;q)_n (y q;q)_n} ,
\end{equation}
while $(y q;q)_\infty$ and $(y q;q)_n$ denote the quantum dilogarithm and the $q$-shifted factorial, respectively. See Appendix~\ref{app: faddeev} for their definitions.
By means of equation~\eqref{chi41}, we can identify
\begin{subequations}
\begin{gather}
 \phi^{({\rm geom}), {\rm WKB}}(X;q) = \log \CJ\bigl(X^2,X;q\bigr) , \label{phi41geom}\\
 \phi^{({\rm conj}), {\rm WKB}}(X;q) = \log \CJ\bigl(X,X^{-1};q\bigr) , \label{phi41conj}
\end{gather}
\end{subequations}
together with $\mathsf{C}_{{\rm geom}} = \mathsf{C}_{{\rm conj}} = \re^{\frac{\pi\ri}{4}}$ and
\begin{subequations}
\begin{gather}
 {\ri s^{({\rm geom})}_0(x) \over \hbar} + s^{({\rm geom})}_1(x) + {\ri s^{({\rm geom})}_0(x_D) \over \hbar_D} + s^{({\rm geom})}_1(x_D) = {x^2 \over\ri \hbar} + {2 \pi x \over\hbar} + x + {\pi^2 \ri \over 3 \hbar} + {\ri \hbar \over 12} , \label{s01geom41}\\
 {\ri s^{({\rm conj})}_0(x) \over \hbar} + s^{({\rm conj})}_1(x) + {\ri s^{({\rm conj})}_0(x_D) \over \hbar_D} + s^{({\rm conj})}_1(x_D) = {\ri x^2 \over 2 \hbar} + { \pi x \over\hbar} + {x \over 2} + {\pi^2 \ri \over 3 \hbar} + {\ri \hbar \over 12} . \label{s01conj41}
\end{gather}
\end{subequations}
Indeed, the two terms in the right-hand side of equation~\eqref{chi41} represent the contributions coming from the geometric and conjugate branches, respectively. These are the only two non-abelian branches of the $\FE$-knot, as described in Example~\ref{41example}.
Moreover, we solve equations~\eqref{s01geom41} and~\eqref{s01conj41} order by order in $x$ and $\hbar$ and obtain that
\begin{subequations}
\begin{alignat}{3}
& s^{({\rm geom})}_0(x) = \frac{\pi^2}{3} - \frac{x^2}{2} , \qquad&& s^{({\rm geom})}_1(x) = x ,& \label{s01geom41-new} \\
& s^{({\rm conj})}_0(x) = \frac{\pi^2}{3} + \frac{x^2}{4} , \qquad&& s^{({\rm conj})}_1(x) = \frac{x}{2} .& \label{s01conj41-new}
\end{alignat}
\end{subequations}
In this way, we have fully determined the perturbative wave functions $\chi^{(\alpha), {\rm WKB}}(u;\hbar)$ of the figure-eight knot in the resummed form of equation~\eqref{WKB-conj}. We verify successfully that they are annihilated by the $\hat A$-operator in equation~\eqref{41Ahat}, as expected.

Let us now test the conjectural integrality structure in equation~\eqref{multi-conj} for both the geometric and the conjugate branches.
Expanding the right-hand side of equations~\eqref{phi41geom} and~\eqref{phi41conj} gives%
\begin{subequations}
\begin{gather}
 \phi^{({\rm geom}), {\rm WKB}}(X;q) = \sum_{k \ge 1} \frac{q^k X^k}{k\bigl(q^k-1\bigr)} + \log \Bigg(\sum_{n \ge 0} \frac{(-1)^n q^{n(n+1)/2} X^{2n}}{\prod_{i=1}^n (1-q^i) (1-q^i X)} \Bigg) , \label{phi41geom-exp}\\
 \phi^{({\rm conj}), {\rm WKB}}(X;q) = \sum_{k \ge 1} \frac{q^k X^{-k}}{k\bigl(q^k-1\bigr)} + \log \Bigg(\sum_{n \ge 0} \frac{ (-1)^n q^{n(n+1)/2} X^{2n}}{\prod_{i=1}^n (1-q^i) (X-q^i)} \Bigg) . \label{phi41conj-exp}
\end{gather}
\end{subequations}
Let us note that both terms on the right-hand side of equation~\eqref{phi41geom-exp} for the geometric branch contribute with positive powers of $X$.
For the conjugate branch, we note that the negative powers of $X$ arise entirely from the first term on the right-hand side of equation~\eqref{phi41conj-exp}, while a~factor of $q^{\sigma_{{\rm conj}, n}}$ with
\be \label{sigma41}
\sigma_{{\rm conj}, n} = \left\lfloor \frac{n^2}{8} \right\rfloor , \qquad n \in \IZ_{>0} ,
\ee
appears in the denominator of the power series in $X$ obtained by expanding the second term. Let us stress that this negative power of $q$ arises from the factors $\prod_{i=1}^n\bigl(X-q^i\bigr)^{-1}$ in the summand inside the logarithm, and it is absent from both the $X^{-1}$-series for the conjugate branch and the $X$-series for the geometric branch.

After extracting the coefficients $a^{(\alpha)}_{\pm n}(q)$, $n \in \IZ_{>0}$, from the series for $\phi^{(\alpha), {\rm WKB}}(X;q)$, one obtains the functions $D_{\pm s}^{(\alpha)}(q)$, $s \in \IZ_{>0}$, by applying the formulae in equations~\eqref{Dtilde1} and~\eqref{Dtilde2}. It is non-trivial that the resulting functions are polynomials with integer coefficients. In particular, the first few of them are
\begin{gather}
D^{({\rm geom})}_1(q) = D^{({\rm geom})}_2(q) = q , \qquad D^{({\rm geom})}_3(q) = q^2 , \qquad D^{({\rm geom})}_4(q)=q^3 , \nonumber\\
D^{({\rm geom})}_5(q) = q^3(1+q) , \qquad D^{({\rm geom})}_6(q) = q^4 (1 + 2 q) ,\nonumber \\
 D^{({\rm geom})}_7(q) = q^4 (1 + q) \bigl(1 + q + q^2\bigr) ,
\end{gather}
for the geometric solution, and
\begin{gather}
D^{({\rm conj})}_1(q) = 0 , \qquad D^{({\rm conj})}_2(q) = D^{({\rm conj})}_3(q) = D^{({\rm conj})}_4(q)=-1 , \qquad D^{({\rm conj})}_5(q) = -1-q ,\nonumber \\
 D^{({\rm conj})}_6(q) = -2-q , \qquad D^{({\rm conj})}_7(q) = -(1 + q) \bigl(1 + q + q^2\bigr) ,
\end{gather}
for its conjugate. Trivially, we have that
\be
D^{({\rm conj})}_{-1}(q) = q , \qquad D^{({\rm conj})}_{-s}(q) = 0 \qquad s \ge 2 .
\ee
We have verified the integrality structure numerically up to $s=20$. It is then straightforward to compute the integers \smash{$c^{(\alpha)}_{\pm s}$}, $s \in \IZ_{>0}$, using equations~\eqref{cs-sequenceP} and~\eqref{cs-sequenceM}, which give
\begin{subequations}
\begin{gather}
\bigl\{ c^{({\rm geom})}_s\bigr\}_{s \in \IZ_{>0}}=\{ 1, 5, 10, 21, 51, 122, 295, 725, 1792, 4455, 11133, 27930, \dots \} , \label{cs41geom} \\
\bigl\{ c^{({\rm conj})}_s\bigr\}_{s \in \IZ_{>0}}= (-1) \{ 0, 4, 9, 20, 50, 121, 294, 724, 1791, 4454, 11132, 27929, \dots \} , \label{cs41conj}
\end{gather}
\end{subequations}
while \smash{$c^{({\rm conj})}_{-s} = 1$} for all $s \in \IZ_{>0}$. We point out that the integers in equation~\eqref{cs41geom} appear to match numerically the sequence\footnote{This is the entry \href{https://oeis.org/A132463}{A132461} in the on-line encyclopedia of integer sequences.}
\be
\label{sequence-ex}
\sum_{k=0}^{\lfloor s/2\rfloor} \left( {s-k \choose k} + {s-k-1 \choose k-1} \right)^2 , \qquad s \in \IZ_{>0} .
\ee
Moreover, we remark that the above sequences of integers satisfy
\be \label{cs-rel41}
c^{({\rm geom})}_s + c^{({\rm conj})}_s = c^{({\rm conj})}_{-s} , \qquad s \in \IZ_{>0} .
\ee
The same integer constants are obtained by implementing the alternative computational method proposed in Appendix~\ref{app: rational2}.

Let us now go back to the quantum $A$-polynomial for the $\FE$-knot written in equation~\eqref{41Ahat}. The corresponding second-order $q$-difference equation is
\be \label{Ahat-41}
\hat{A}_{\FE}\bigl(\hat{P}, \hat{X}, q\bigr) \chi^{(\alpha), {\rm WKB}}(u; \hbar) = 0 ,
\ee
where \smash{$\chi^{(\alpha), {\rm WKB}}(u; \hbar)$} is the resummed perturbative wave function in equation~\eqref{chi-matrix-2} associated with the branch $\alpha$.
In the rational case of \smash{$q= \re^{2 \pi \ri {P \over Q}}$} with $P, Q \in \IZ_{>0}$ coprime, equation~\eqref{Ahat-41} can be solved in closed form by directly applying the results of Section~\ref{sec: 2nd-order}.
For simplicity, let us consider the classical solution $\alpha = {\rm geom}$ and drop the label from our notation.
After suitably taking into account the exponential prefactor containing the functions $s_{0,1}(x)$ in equation~\eqref{s01geom41-new}, the $q$-difference equation above assumes the form in equation~\eqref{Psi-eq} with the identifications%
 \begin{subequations} \label{coeffs-qdiff}
 \begin{align}
 A(X) &= q^{-2} X^{-2} \frac{C_1(X,q)}{C_2(X,q)} = -q^{-2} X^{-2} {X^2-1 \over q^{-1}X^2-1} \tilde{a}(X;q) , \label{coeffs-qdiffA} \\
 B(X) &= q^{-2} X^{-4} \frac{C_0(X,q)}{C_2(X,q)}= q^{-2} X^{-4} {qX^2-1 \over q^{-1}X^2-1} , \label{coeffs-qdiffB}
 \end{align}
 \end{subequations}
 where $C_0$, $C_1$, $C_2$ are the coefficient functions defined in equation~\eqref{41xC} and we have introduced
 \begin{gather} \label{atilde}
 \tilde{a}(X;q) = X^2+ X^{-2} -X-X^{-1}-q-q^{-1} .
 \end{gather}
Substituting the expressions in equations~\eqref{coeffs-qdiffA} and~\eqref{coeffs-qdiffB} into equation~\eqref{matrix2}, we compute the matrix $\CM_Q(X)$ and obtain explicit formulae for $\operatorname{tr} \CM_Q(X)$, $\det \CM_Q(X)$, and $\Delta_Q(X)$. Specifically, we find that
\begin{subequations} \label{funct-mat41}
\begin{gather}
\operatorname{tr} \CM_Q(X) = X^{-4 Q} \bigl(1- X^Q - 2 X^{2 Q} - X^{3 Q} + X^{4 Q} \bigr) , \\
\det \CM_Q(X) =X^{-4 Q} , \\
\Delta_Q(X) = X^{-8 Q}\bigl(1-X^{2 Q}\bigr)^2 \bigl(1-2 X^{Q} -X^{2 Q}-2 X^{3 Q}+X^{4 Q}\bigr) ,
\end{gather}
\end{subequations}
and plugging these into equation~\eqref{2nd-end} with a choice of plus sign to satisfy the requirement in equation~\eqref{branch} for $\alpha={\rm geom}$, we obtain the functions $\Upsilon (X)$ and $\varphi_Q(X)$. Observe that all functions in equation~\eqref{funct-mat41}, and thus also $\varphi_Q(X)$, depend on $X$ through $X^Q$, as expected, and we can identify
\begin{subequations}
\begin{gather}
\operatorname{tr} \CM_Q(X) = X^{-4 Q} T\bigl(X^Q\bigr) = X^{-2 Q} \tilde{a}\bigl(X^Q; 1\bigr) , \\
\Delta_Q(X) = X^{-8 Q}\bigl(1-X^{2 Q}\bigr)^2 \Delta\bigl(X^Q\bigr) = X^{-4 Q} \bigl(\tilde{a}\bigl(X^Q; 1\bigr)^2 - 4 \bigr) ,
\end{gather}
\end{subequations}
where $\tilde{a}(X; q)$ is defined in equation~\eqref{atilde}, while $T(X)$ and $\Delta(X)$ are the functions introduced in equations~\eqref{classicT} and~\eqref{classicD}, respectively, which appear in the formula for the solutions to the classical $A$-polynomial of the figure-eight knot in equation~\eqref{classicP}.
In particular, it follows that
\begin{align}
\varphi_Q(X)&= \frac{1}{Q} \log \left(\frac{T\bigl(X^Q\bigr) + \bigl(1-X^{2 Q}\bigr)\sqrt{\Delta\bigl(X^Q\bigr)}}{2 X^{4 Q}} \right)\nonumber\\
& = \frac{1}{Q} \log \left(\frac{\tilde{a}\bigl(X^Q; 1\bigr) + \sqrt{\tilde{a}\bigl(X^Q; 1\bigr)^2-4}}{2 X^{2 Q}} \right) , \label{8varphiQ}
\end{align}
which gives us another way to numerically evaluate the integer constants $c_s$, $s \in \IZ_{>0}$, for the geometric branch, following equation~\eqref{varphiQ-cs} for any choice of $P, Q \in \IZ_{>0}$ coprime. This is in full agreement with our previous computations. In principle, one could use the formula for $\varphi_Q(X)$ with $P=Q=1$, that is,
\be
\label{8varphi1}
\varphi_1(X)= \log \left( {\tilde{a}(X;1) + {\sqrt{\tilde{a}(X;1)^2- 4}} \over 2 X^2} \right) ,
\ee
to derive the conjectural expression in equation~\eqref{sequence-ex}, but we have not attempted to do so. The same formalism can be applied to the conjugate branch of the figure-eight knot, in which case the appropriate functions $s_{0,1}(x)$ are written in equation~\eqref{s01conj41-new} and the sign ambiguity in equation~\eqref{2nd-end} is resolved by imposing the condition in equation~\eqref{branch} with $\alpha = \mathrm{conj}$, which implies a choice of minus sign. Again, the corresponding computation of the integer constants~$c_{\pm s}$,~$s \in \IZ_{>0}$, for the conjugate branch agrees with the results obtained above.

\subsection{The three-twist knot} \label{sec: 52}
The state integral of the three-twist knot $\chi_{{\bf{5}}_2}(u;\hbar)$ can be expressed explicitly as~\cite{bdp,dimofte-levelk,ggm2}
 \begin{align}
 \chi_{{\bf{5}}_2}(u;\hbar)={}&\re^{-\frac{\pi\ri}{12}-\frac{5\pi\ri}{3}c_\mb^2}\bigl(
 \re^{\pi\ri(-2u^2-4c_\mb u)}
 \CH\bigl(\re^{2 \pi \mb u},\re^{4 \pi \mb u},\re^{4 \pi \mb u}, ;q\bigr)\nonumber\\
 &\times
 \CH\bigl(\re^{2 \pi \mb^{-1} u},\re^{4 \pi \mb^{-1} u},\re^{4 \pi \mb^{-1} u};\tq^{-1}\bigr)\nonumber\\
 &\phantom{\times}{}+\re^{\pi\ri(-2u^2+4c_\mb u)}
 \CH\bigl(\re^{-2 \pi \mb u},\re^{-4 \pi \mb u},\re^{-4 \pi \mb u};q\bigr)\nonumber\\
 &\times \CH\bigl(\re^{-2 \pi \mb^{-1} u},\re^{-4 \pi \mb^{-1} u},\re^{-4 \pi \mb^{-1} u};\tq^{-1}\bigr) \nonumber\\
 &\phantom{\times}{}+\CH\bigl(\re^{2 \pi \mb u},\re^{-2 \pi \mb u},1;q\bigr)
 \CH\bigl(\re^{2 \pi \mb^{-1} u},\re^{-2 \pi \mb^{-1} u},1;\tq^{-1}\bigr)
 \bigr) ,\label{chi52}
\end{align}
where $c_{\mb}= \ri \bigl(\mb + \mb^{-1}\bigr)/2$ and the $q$-special function $\CH(x,y,z;q)$ is defined by\footnote{$\CH(x,y;q)$ is closely related to the $q$-hypergeometric function.}
 \begin{equation} \label{H}
 \CH(x,y,z;q) = (x q;q)_\infty (y q;q)_\infty \sum_{n=0}^\infty \frac{q^{n(n+1)} z^n}{(q;q)_n (x q;q)_n (y q;q)_n} ,
 \end{equation}
while $(x q;q)_\infty$ and $(x q;q)_n$ denote the quantum dilogarithm and the $q$-shifted factorial, respectively. See Appendix~\ref{app: faddeev} for their definitions.
As for the case of the figure-eight knot in Section~\ref{sec: 41}, using equation~\eqref{chi52}, we can identify
\begin{subequations}
\begin{gather}
 \phi^{({\rm geom}), {\rm WKB}}(X;q) = \log \CH\bigl(X,X^2,X^2;q\bigr) , \label{phi52geom} \\
 \phi^{({\rm conj}), {\rm WKB}}(X;q) = \log \CH\bigl(X^{-1},X^{-2},X^{-2};q\bigr) , \label{phi52conj} \\
 \phi^{({\rm self}), {\rm WKB}}(X;q) = \log \CH\bigl(X,X^{-1},1;q\bigr) , \label{phi52self}
\end{gather}
\end{subequations}
together with \smash{$\mathsf{C}_{{\rm geom}} = \mathsf{C}_{{\rm conj}} = \mathsf{C}_{{\rm self}} = \re^{\frac{3 \pi \ri}{4}}$} and
\begin{subequations}
\begin{gather}
 {\ri s^{({\rm geom})}_0(x) \over \hbar}+ s^{({\rm geom})}_1(x) + {\ri s^{({\rm geom})}_0(x_D) \over \hbar_D} + s^{({\rm geom})}_1(x_D)
 \nonumber\\
 \qquad= {x^2 \over\ri \hbar} + {2 \pi x \over\hbar} + x +{5 \pi^2 \ri \over 6 \hbar} + {5 \ri \hbar \over 24} , \label{s01geom52}\\
 {\ri s^{({\rm conj})}_0(x) \over \hbar} + s^{({\rm conj})}_1(x) + {\ri s^{({\rm conj})}_0(x_D) \over \hbar_D} + s^{({\rm conj})}_1(x_D)
 = {x^2 \over \ri \hbar} - { 2 \pi x \over\hbar}- x + {5 \pi^2 \ri \over 6 \hbar} + {5 \ri \hbar \over 24} , \label{s01conj52}\\
 {\ri s^{({\rm self})}_0(x) \over \hbar} + s^{({\rm self})}_1(x) + {\ri s^{({\rm self})}_0(x_D) \over \hbar_D} + s^{({\rm self})}_1(x_D)
 = {5 \pi^2 \ri \over 6 \hbar} + {5 \ri \hbar \over 24} .\label{s01self52}
\end{gather}
\end{subequations}
Indeed, the three terms in the right-hand side of equation~\eqref{chi52} capture the contributions to the exact partition function coming from the geometric, conjugate, and self-conjugate branches, respectively. These are the only three non-abelian branches of the $\TT$-knot, as described in Example~\ref{52example}.
Moreover, as before, we solve equations~\eqref{s01geom52},~\eqref{s01conj52}, and~\eqref{s01self52} order by order in $x$ and $\hbar$ and obtain that
\begin{subequations}
\begin{gather}
 s^{({\rm geom})}_0(x) = \frac{5 \pi^2}{6} - \frac{x^2}{2} , \qquad s^{({\rm geom})}_1(x) = x , \label{s01geom52-new}\\
 s^{({\rm conj})}_0(x) = \frac{5 \pi^2}{6} - \frac{x^2}{2} , \qquad s^{({\rm conj})}_1(x) = -x , \label{s01conj52-new}\\
 s^{({\rm self})}_0(x) = \frac{5 \pi^2}{6} , \qquad s^{({\rm self})}_1(x) = 0 . \label{s01self52-new}
\end{gather}
\end{subequations}
In this way, we have fully determined the perturbative wave functions $\chi^{(\alpha), {\rm WKB}}(u;\hbar)$ of the three-twist knot in the resummed form of equation~\eqref{WKB-conj}. As before, we verify successfully that they are annihilated by the $\hat A$-operator in equation~\eqref{52Ahat}, as expected.

Let us now test the integrality conjecture in equation~\eqref{multi-conj} for all three non-abelian branches.
Expanding the right-hand side of equations~\eqref{phi52geom},~\eqref{phi52conj}, and~\eqref{phi52self} gives
\begin{subequations}
\begin{gather}
 \phi^{({\rm geom}), {\rm WKB}}(X;q)= \sum_{k \ge 1} \frac{q^k \bigl(X^k+X^{2k}\bigr)}{k\bigl(q^k-1\bigr)}\nonumber \\
 \phantom{ \phi^{({\rm geom}), {\rm WKB}}(X;q)=}{}+ \log \left(\sum_{n \ge 0} \frac{q^{n(n+1)} X^{2n}}{\prod_{i=1}^n \bigl(1-q^i\bigr) \bigl(1-q^i X\bigr)\bigl(1-q^i X^2\bigr)} \right) , \label{phi52geom-exp}\\
 \phi^{({\rm conj}), {\rm WKB}}(X;q) = \sum_{k \ge 1} \frac{q^k \bigl(X^{-k}+X^{-2k}\bigr)}{k\bigl(q^k-1\bigr)} \nonumber\\
 \phantom{ \phi^{({\rm conj}), {\rm WKB}}(X;q) =}{}+ \log \left(\sum_{n \ge 0} \frac{q^{n(n+1)} X^{n}}{\prod_{i=1}^n \bigl(1-q^i\bigr) \bigl(X-q^i\bigr)\bigl(X^2-q^i\bigr)} \right) , \label{phi52conj-exp}\\
 \phi^{({\rm self}), {\rm WKB}}(X;q) = \sum_{k \ge 1} \frac{q^k \bigl(X^k+X^{-k}\bigr)}{k\bigl(q^k-1\bigr)}\nonumber\\
 \phantom{ \phi^{({\rm self}), {\rm WKB}}(X;q) = }{}+ \log \left(\sum_{n \ge 0} \frac{q^{n(n+1)} X^{n}}{\prod_{i=1}^n \bigl(1-q^i\bigr) \bigl(1-q^i X\bigr)\bigl(X-q^i\bigr)} \right) . \label{phi52self-exp}
\end{gather}
\end{subequations}
Both terms on the right-hand side of equation~\eqref{phi52geom-exp} for the geometric branch contribute with positive powers of $X$.
For the conjugate branch, we note that the negative powers of $X$ arise entirely from the first term on the right-hand side of equation~\eqref{phi52conj-exp}, while a factor of $q^{\sigma_{{\rm conj}, n}}$ with
\be \label{sigma52}
\sigma_{{\rm conj}, n} = \left\lfloor \frac{n^2}{4} \right\rfloor , \qquad n \in \IZ_{>0} ,
\ee
appears in the denominator of the power series in $X$ obtained by expanding
the second term. Let us stress that this negative power of $q$ arises from the
factors\footnote{Note that there are twice as many contributing factors of $q^{-1}$ here compared to the conjugate branch of the figure-eight knot, in agreement with the observed formulae for the functions $\sigma_{{\rm conj}, n}$ in equations~\eqref{sigma41} and~\eqref{sigma52}.} $\prod_{i=1}^n\bigl(X-q^i\bigr)^{-1}\bigl(X^2-q^i\bigr)^{-1}$ in the summand inside the logarithm, and it is absent from both the $X^{-1}$-series for the conjugate branch and the $X$-series
for the geometric branch.
Finally, in the case of the self-conjugate branch, the first term on the right-hand side
of equation~\eqref{phi52self-exp} supplies both positive and negative powers of $X$,
while the second term only adds to the $X$-series after expansion. Hence, a more
complicated factor of $q^{\sigma_{{\rm self}, n}}$ with
\be \label{sigma52self}
\sigma_{{\rm self}, n} = \left\lfloor \frac{1}{24} \left(P \left(\left\lfloor \frac{2 (n+2)}{3}
\right\rfloor \right)^2-1 \right) \right\rfloor , \qquad n \in \IZ_{>0} ,
\ee
where $P(n)$ denotes the $n$-th prime number, occurs in the denominator of the power series in~$X$ after summing up the contribution from the expansion of the second term on the right-hand side of equation~\eqref{phi52self-exp}.
This originates from the factors $\prod_{i=1}^n\bigl(X-q^i\bigr)^{-1}$ in the summand inside the logarithm.

As we have done for the $\FE$-knot, after extracting the coefficients \smash{$a^{(\alpha)}_{\pm n}(q)$}, $n \in \IZ_{>0}$, from the series for \smash{$\phi^{(\alpha), {\rm WKB}}(X;q)$}, we obtain the functions \smash{$D_{\pm s}^{(\alpha)}(q)$}, $s \in \IZ_{>0}$, by applying the formulae in equations~\eqref{Dtilde1} and~\eqref{Dtilde2}. Again, these are non-trivial polynomials with integer coefficients. In particular, the first few of them are
\begin{gather}
D^{({\rm geom})}_1(q) = q , \qquad D^{({\rm geom})}_2(q) = -q(-1 + q) , \qquad D^{({\rm geom})}_3(q) = D^{({\rm geom})}_4(q)=-q^3 , \nonumber\\
D^{({\rm geom})}_5(q) = q^4 \bigl(-1 + q^2\bigr) , \qquad D^{({\rm geom})}_6(q) = q^4 \bigl(-1 + 2 q^2 + q^4\bigr) , \nonumber\\
D^{({\rm geom})}_7(q) = q^5 \bigl(-1 + q (1 + q)^2\bigr) ,
\end{gather}
for the geometric solution,
\begin{gather}
D^{({\rm conj})}_1(q) = D^{({\rm conj})}_2(q) = -1 , \qquad D^{({\rm conj})}_3(q) = -2 - q , \qquad D^{({\rm conj})}_4(q)=-(1 + q) (1 + 2 q) ,\nonumber \\
D^{({\rm conj})}_5(q) = -2 - 2 q - 7 q^2 - 4 q^3 - q^4 , \nonumber\\
D^{({\rm conj})}_6(q) = -(1 + q) \bigl(1 + 2 q + 3 q^2 + 6 q^3 + 7 q^4 + 2 q^5\bigr), \nonumber\\
D^{({\rm conj})}_7(q) = -2 - 2 q - 8 q^2 - 11 q^3 - 20 q^4 - 23 q^5 - 32 q^6 - 19 q^7 - 6 q^8 - q^9 ,
 \end{gather}
for its conjugate, and
\begin{gather}
D^{({\rm self})}_1(q) = 2 q , \qquad D^{({\rm self})}_2(q) = 1 + q^2 , \qquad D^{({\rm self})}_3(q) =1 + q^2 + 2 q^4 , \nonumber\\
D^{({\rm self})}_4(q) =-1 - q + q^2 + 2 q^4 + 3 q^6 + q^7 ,\nonumber \\
D^{({\rm self})}_5(q) = -1 + q - q^2 + 2 q^3 + 4 q^5 + q^6 + 6 q^7 + 2 q^8 + 2 q^9 ,\nonumber \\
D^{({\rm self})}_6(q) = -1 - q^2 - 2 q^4 + 2 q^5 - 2 q^6 + 4 q^7 + q^8 + 8 q^9 + 4 q^{10}\nonumber \\
\phantom{D^{({\rm self})}_6(q) =}{} + 11 q^{11} + 7 q^{12} + 5 q^{13} + 3 q^{14} + q^{15} ,\nonumber \\
D^{({\rm self})}_7(q) = -1 - q^3 - 2 q^5 + q^6 - 4 q^7 + 4 q^8 - 3 q^9 + 9 q^{10} + 4 q^{11} + 18 q^{12} \nonumber\\
\phantom{D^{({\rm self})}_7(q) =}{}+ 12 q^{13} + 27 q^{14} + 18 q^{15} + 17 q^{16} + 10 q^{17} + 8 q^{18} +
 2 q^{19} + 2 q^{20} ,
\end{gather}
for the self-conjugate one. Trivially, we have that
\begin{gather}
D^{({\rm conj})}_{-1}(q) = D^{({\rm conj})}_{-2}(q) = q , \qquad D^{({\rm conj})}_{-s}(q) = 0, \qquad s \ge 3 , \nonumber\\
D^{({\rm self})}_{-1}(q) = q , \qquad D^{({\rm self})}_{-s}(q) = 0, \qquad s \ge 2 .
\end{gather}
We have verified numerically that the polynomials $D_{\pm s}(q)$ have integer coefficients up to $s=20$ for all three branches. We then compute the integers \smash{$c^{(\alpha)}_{\pm s}$}, $s \in \IZ_{>0}$, using equations~\eqref{cs-sequenceP} and~\eqref{cs-sequenceM}, which give
\begin{subequations}
\begin{gather}
\big\{ c^{({\rm geom})}_s\big\}_{s \in \IZ_{>0}}=\{ 1, 1, -8, -15, 1, 64, 148, 49, -575, -1599, -1088, 5088, \dots \} , \label{cs52geom} \\
\big\{ c^{({\rm conj})}_s\big\}_{s \in \IZ_{>0}}= (-1) \{ 1, 5, 28, 101, 401, 1544, 6077, 24101, 96418, 388205, 1571307,\nonumber\\
\phantom{\big\{ c^{({\rm conj})}_s\big\}_{s \in \IZ_{>0}}= }{} 6387608, \dots \} , \label{cs52conj}\\
\big\{ c^{({\rm self})}_s\big\}_{s \in \IZ_{>0}}= \{ 2, 10, 38, 122, 402, 1486, 5931, 24058, 96995, 389810, 1572397,\nonumber\\
\phantom{\big\{ c^{({\rm self})}_s\big\}_{s \in \IZ_{>0}}=}{} 6382526, \dots \} , \label{cs52self}
\end{gather}
\end{subequations}
while \smash{$c^{({\rm conj})}_{-s} = [s]_2+ 5 [s+1]_2$}, where $[s]_2$ denotes the remainder of the division of $s$ by $2$, and~\smash{$c^{({\rm self})}_{-s} = 1$} for all $s \in \IZ_{>0}$.
We observe that the above sequences of integers satisfy the relation
\be \label{cs-rel52}
c^{({\rm geom})}_s + c^{({\rm conj})}_s + c^{({\rm self})}_s = c^{({\rm conj})}_{-s} + c^{({\rm self})}_{-s} , \qquad s \in \IZ_{>0} ,
\ee
corresponding to the analogous formula in equation~\eqref{cs-rel41} for the figure-eight knot. Note that, in both our examples, the sum over all non-abelian branches of the integers \smash{$c^{(\alpha)}_s$} at fixed $s$ equals the sum of the integers \smash{$c^{(\alpha)}_{-s}$}, that is,
\be
\sum_{\alpha} c^{(\alpha)}_{s} = \sum_{\alpha} c^{(\alpha)}_{-s} , \qquad s \in \IZ_{>0} ,
\ee
which appears to be a periodic function of $s$.
Again, the alternative computational approach of Appendix~\ref{app: rational2} agrees with the above results.

Let us now go back to the quantum $A$-polynomial for the $\TT$-knot written in equation~\eqref{52Ahat}. The corresponding $q$-difference equation in the form of equation~\eqref{AJ-eq} is of third order. Precisely, we have
\be \label{Ahat-52}
\hat{A}_{\TT}\bigl(\hat{P}, \hat{X}, q\bigr) \chi^{(\alpha), {\rm WKB}}(u; \hbar) = 0 ,
\ee
where $\chi^{(\alpha), {\rm WKB}}(u; \hbar)$ is the resummed perturbative wave function in equation~\eqref{chi-matrix-2} associated with the branch $\alpha$.
In the rational case of \smash{$q= \re^{2 \pi \ri {P \over Q}}$} with $P, Q \in \IZ_{>0}$ coprime, equation~\eqref{Ahat-52} can be solved in closed form by directly applying the results of Section~\ref{sec: 3rd-order}.
As for the $\FE$-knot, let us consider the case $\alpha = {\rm geom}$ and drop the label to simplify our notation.
After appropriately accounting for the exponential prefactor containing the functions $s_{0,1}(x)$ in equation~\eqref{s01geom52-new}, the $q$-difference equation above assumes the form in equation~\eqref{Psi-eq3} with the identifications
 \begin{gather}
 \begin{split}
& A(X)= q^{-2} X^{-2} \frac{C_2(X,q)}{C_3(X,q)} , \qquad B(X)= q^{-2} X^{-4} \frac{C_1(X,q)}{C_3(X,q)} , \\
& C(X) = X^{-6} \frac{C_0(X,q)}{C_3(X,q)} ,\end{split}
\label{coeffs-qdiff52}
 \end{gather}
 where $C_0$, $C_1$, $C_2$, $C_3$ are the coefficient functions defined in equation~\eqref{52xC}. Substituting the expressions in equation~\eqref{coeffs-qdiff52} into equation~\eqref{matrix3} and computing the matrix $\CM_Q(X)$, we can then derive explicit formulae for $\Pi \u$ and $\u_0$ from equations~\eqref{Pi3} and~\eqref{u03}. We stress that the ambiguity in the definition of $\Theta_2$ is resolved by requiring equation~\eqref{branch} to be satisfied with~${\alpha={\rm geom}}$, which implies the choice $m=2$. Plugging these into equation~\eqref{dict-end}, we find the functions $\Upsilon (X)$ and $\varphi_Q(X)$ for the geometric branch of the three-twist knot. We can easily use the resulting explicit expression for $\varphi_Q(X)$ to numerically evaluate the integer constants $c_s$, $s \in \IZ_{>0}$, for the geometric branch by applying equation~\eqref{varphiQ-cs} for any choice of $P, Q \in \IZ_{>0}$ coprime, thus producing a third independent computation which is in perfect agreement with the previous two. The same discussion can be applied to the conjugate and self-conjugate branches of the $\TT$-knot, in which case the appropriate functions $s_{0,1}(x)$ are written in equations~\eqref{s01conj52-new} and~\eqref{s01self52-new} and the ambiguity in the choice of $m$ in equation~\eqref{theta2-m} is fixed by imposing the condition in equation~\eqref{branch} with $\alpha = \mathrm{conj}, \mathrm{self}$, yielding $m=1$ and $m=0$, respectively. Again, the corresponding integer constants $c_{\pm s}$, $s \in \IZ_{>0}$, for the conjugate and self-conjugate branches match our previous computations.

\subsection{Proofs of admissibility} \label{sec: proofs-examples}
As mentioned in Section~\ref{sec: conjectures}, the conjectured integrality property of the resummed perturbative wave function in equation~\eqref{WKB-conj} can be stated as admissibility in the sense of Kontsevich--Soibelman.

Recall that a formal power series in one variable $z$ with coefficients in the ring of formal Laurent series $\IZ\bigl(\!\bigl(q^{1/2}\bigr)\!\bigr)$, that is, $F(z; q) = \sum_{n \ge 0} A_n z^n \in \IZ\bigl(\!\bigl(q^{1/2}\bigr)\!\bigr)[\![z]\!]$, is admissible if and only if it can be expressed as a (possibly infinite) product of quantum dilogarithms of the form
\be \label{admissible-1}
F(z;q)=\prod_{s \ge 1} \prod_{i \in \IZ} \bigl(q^{i/2} z^s; q\bigr)_\infty^{c(s,i)} \in 1 + z \IZ\bigl(\!\bigl(q^{1/2}\bigr)\!\bigr)[\![z]\!] , \qquad c(s,i) \in \IZ ,
\ee
where $c(s,i)=0$ for $|i| \gg 0$ and fixed $s$. It follows that admissible series form a group under multiplication.
Note that the formula above is equivalent to
\be \label{admissible-2}
F(z;q)=\exp \left(\sum_{k,s \ge 1} \frac{f_s\bigl(q^{k/2}\bigr)}{k \bigl(q^k-1\bigr)} z^{s k} \right) , \qquad f_s(q) \in \IZ\big[q^\pm\big] ,
\ee
which reproduces in $z$ the same integrality structure we conjectured for the $X$, $X^{-1}$-series in equation~\eqref{multi-conj} with one caveat. More precisely, for all choices of non-abelian branch $\alpha$ and up to the exponential prefactors, the resummed WKB wave function in equation~\eqref{WKB-conj} is expected to be a product of two admissible series in $X$ and $X^{-1}$, respectively, such that $q^{-k \sigma_s} D_s\bigl(q^k\bigr)= f_s\bigl(q^{k/2}\bigr) \in \IZ\big[q^\pm\big]$ for all $k$, $s$.

The notion of admissibility can be appropriately extended\footnote{In fact, admissibility is defined in~\cite{ks} for series in $R\bigl(\!\bigl(q^{1/2}\bigr)\!\bigr)[\![\{z_i\}_{i \in I}]\!]$, where $R$ is a $\lambda$-ring and $q^{1/2}, \{z_i\}_{i \in I}$ are line elements.} to apply to multi-variable series in $\IZ\bigl(\!\bigl(q^{1/2}\bigr)\!\bigr)[\![\{z_i\}_{i \in I}]\!]$, where $I={1, \dots, |I|}$.
The authors of~\cite{ks} proved that admissibility is preserved under certain transformations. In particular, for any given symmetric integer matrix $B=(b_{i,j})_{i,j=1,\dots,|I|}$, a series
\be
F(z;q)= \sum_{\{n_i\} \in \IZ_{\ge 0}^{|I|}} A_n \prod_{i \in I} z_i^{n_i} \in \IZ\bigl(\!\bigl(q^{1/2}\bigr)\!\bigr)[\![\{z_i\}_{i \in I}]\!]
\ee
is admissible if and only if the twisted series
\be \label{admissible-3}
\tilde{F}(z;q)= \sum_{\{n_i\} \in \IZ_{\ge 0}^{|I|}} \bigl(-q^{1/2}\bigr)^{\sum_{i,j=1}^{|I|} b_{i,j} n_i n_j} A_n \prod_{i \in I} z_i^{n_i}
\ee
is admissible~\cite[Theorem~9]{ks}. In what follows, we will refer to this statement as the Kontsevich--Soibelman (KS) theorem.

We will now show that the $q$-special functions appearing as $\exp\bigl(\phi^{(\alpha), {\rm WKB}}(X;q)\bigr)$ in the holomorphic/antiholomorphic block decompositions in equations~\eqref{chi41} and~\eqref{chi52} for the non-perturbative wave functions of the figure-eight and three-twist knots are admissible.\footnote{We thank the anonymous referee for providing us with a sketch of the proof.}

\begin{proof}[Proof of admissibility for the $\FE$-knot.]
Let us consider each classical branch separately.

(1) For $\alpha = {\rm geom}$, equations~\eqref{J} and~\eqref{phi41geom} imply the $q$-series representation
\be \label{fact-admissible-1}
\ba
\re^{\phi^{({\rm geom}), {\rm WKB}}(X;q)} &= (qX;q)_\infty \sum_{n=0}^\infty \frac{(-1)^n q^{n(n+1)/2}}{(q;q)_n (qX;q)_n} X^{2n} \\
&= \sum_{m,n =0}^\infty q^{nm} \frac{(-1)^n q^{n(n+1)/2}}{(q;q)_n} X^{2n} \frac{(-1)^m q^{m(m+1)/2}}{(q;q)_m} X^{m} ,
\ea
\ee
where we have applied the identities in equations~\eqref{eq: qFactor2} and~\eqref{eq: qID-1} in the second equality. Observe that the twist factor $q^{nm}$ corresponds to the symmetric integer matrix
\be
B_{\FE} = \begin{pmatrix}
		0 & 1 \\
		1 & 0
	\end{pmatrix} ,
\ee
while the untwisted series
\be
\sum_{m,n =0}^\infty \frac{(-1)^n q^{n(n+1)/2}}{(q;q)_n} X^{2n} \frac{(-1)^m q^{m(m+1)/2}}{(q;q)_m} X^{m} = \bigl(qX^2;q\bigr)_\infty (qX;q)_\infty
\ee
is the product of two quantum dilogarithms and therefore admissible. We conclude by means of the KS theorem.

(2) For $\alpha = {\rm conj}$, equations~\eqref{J} and~\eqref{phi41conj} imply the $q$-series representation
\begin{align}
\re^{\phi^{({\rm conj}), {\rm WKB}}(X;q)} &= \bigl(qX^{-1};q\bigr)_\infty \sum_{n=0}^\infty \frac{(-1)^n q^{n(n+1)/2}}{(q;q)_n \bigl(qX^{-1};q\bigr)_n} X^n \nonumber\\
&= \theta(X; q) \sum_{m,n =0}^\infty q^{-nm} \frac{X^{2n}}{(q;q)_n} \frac{X^{m}}{(q;q)_m} ,\label{fact-admissible-2}
\end{align}
where we have applied the identities in equations~\eqref{eq: qFactorM},~\eqref{eq: qFactor2}, and~\eqref{eq: qID-2} in the second equality, and we have denoted by
\be \label{q-theta}
\theta(X; q) = \bigl(qX^{-1};q\bigr)_\infty (X;q)_\infty
\ee
the $q$-theta function.
Observe that the twist factor $q^{-nm}$ corresponds to the symmetric integer matrix
\be
-B_{\FE} = \begin{pmatrix}
		0 & -1 \\
		-1 & 0
	\end{pmatrix} ,
\ee
while the untwisted series
\be
\sum_{m,n =0}^\infty \frac{X^{2n}}{(q;q)_n} \frac{X^{m}}{(q;q)_m} = \frac{1}{\bigl(X^2;q\bigr)_\infty (X;q)_\infty}
\ee
is the product of two quantum dilogarithms and therefore admissible. Again, we conclude by means of the KS theorem.
\end{proof}

\begin{proof}[Proof of admissibility for the $\TT$-knot.]
Let us consider each classical branch separately.

(1) For $\alpha = {\rm geom}$, equations~\eqref{H} and~\eqref{phi52geom} imply that
\begin{gather}
\re^{\phi^{({\rm geom}), {\rm WKB}}(X;q)}\nonumber \\
\qquad= (qX;q)_\infty \bigl(qX^2;q\bigr)_\infty \sum_{n=0}^\infty \frac{q^{n(n+1)}}{(q;q)_n (qX;q)_n \bigl(qX^2;q\bigr)_n} X^{2n} \nonumber\\
\qquad= \sum_{m,n,k =0}^\infty q^{nm+nk} \frac{q^{n(n+1)}}{(q;q)_n} X^{2n} \frac{(-1)^m q^{m(m+1)/2}}{(q;q)_m} X^{m} \frac{(-1)^k q^{k(k+1)/2}}{(q;q)_k} X^{2k} ,\label{fact-admissible-3}
\end{gather}
where we have used the identities in equations~\eqref{eq: qFactor2} and~\eqref{eq: qID-1} in the second step. Observe that the twist factor $q^{nm+nk}$ corresponds to the symmetric integer matrix
\be
B_{\TT} = \begin{pmatrix}
		0 & 1 & 1 \\
		1 & 0 & 0 \\
		1 & 0 & 0 \\
	\end{pmatrix} ,
\ee
while the untwisted series is the product of three admissible series\footnote{The third factor in equation~\eqref{3fact-twist} can be thought of as the twist of $1/(qX;q)_\infty$ by a factor \smash{$q^{n^2}$}, which gives a one-dimensional matrix $B=(2)$.}{\samepage
\be \label{3fact-twist}
\bigl(qX^2;q\bigr)_\infty (qX;q)_\infty \sum_{n =0}^\infty \frac{q^{n(n+1)}}{(q;q)_n} X^{2n} .
\ee
The KS theorem applies.}

(2) For $\alpha = {\rm conj}$, equations~\eqref{H} and~\eqref{phi52conj} imply that
\be \label{fact-admissible-4}
\ba
\re^{\phi^{({\rm conj}), {\rm WKB}}(X;q)} &= \bigl(qX^{-1};q\bigr)_\infty \bigl(qX^{-2};q\bigr)_\infty \sum_{n=0}^\infty \frac{q^{n(n+1)}}{(q;q)_n \bigl(qX^{-1};q\bigr)_n \bigl(qX^{-2};q\bigr)_n} X^{-2n} \\
&= \theta(X; q) \theta\bigl(X^2; q\bigr) \sum_{m,n,k =0}^\infty q^{-nm-nk} \frac{X^{2n}}{(q;q)_n} \frac{X^{m}}{(q;q)_m} \frac{X^{2k}}{(q;q)_k} ,
\ea
\ee
where we have used the identities in equations~\eqref{eq: qFactorM},~\eqref{eq: qFactor2}, and~\eqref{eq: qID-2} in the second step. The $q$-theta function $\theta(X; q)$ is introduced in equation~\eqref{q-theta}. Observe that the twist factor $q^{-nm-nk}$ corresponds to the symmetric integer matrix
\be
-B_{\TT} = \begin{pmatrix}
		0 & -1 & -1 \\
		-1 & 0 & 0 \\
		-1 & 0 & 0 \\
	\end{pmatrix} ,
\ee
while the untwisted series is the product of three quantum dilogarithms
\be
\frac{1}{\bigl(X^2;q\bigr)^2_\infty (X;q)_\infty} .
\ee
The KS theorem applies.

(3) For $\alpha = {\rm self}$, equations~\eqref{H} and~\eqref{phi52self} imply that
\begin{align}
\re^{\phi^{({\rm self}), {\rm WKB}}(X;q)} &= (qX;q)_\infty \bigl(qX^{-1};q\bigr)_\infty \sum_{n=0}^\infty \frac{q^{n(n+1)}}{(q;q)_n (qX;q)_n \bigl(qX^{-1};q\bigr)_n} \nonumber\\
&= \theta(X; q) \sum_{m,n,k =0}^\infty q^{nm-nk} \frac{(-1)^n q^{n(n+1)/2}}{(q;q)_n} \frac{(-1)^m q^{m(m+1)/2}}{(q;q)_m} X^m \frac{X^{k}}{(q;q)_k} ,\label{fact-admissible-5}
\end{align}
where we have used the identities in equations~\eqref{eq: qFactorM},~\eqref{eq: qFactor2},~\eqref{eq: qID-1}, and~\eqref{eq: qID-2} in the second step. The $q$-theta function $\theta(X; q)$ is introduced in equation~\eqref{q-theta}. Observe that the twist factor $q^{nm-nk}$ corresponds to the symmetric integer matrix
\be
B_{\TT}' = \begin{pmatrix}
		0 & 1 & -1 \\
		1 & 0 & 0 \\
		-1 & 0 & 0 \\
	\end{pmatrix} ,
\ee
while the untwisted series is the product of three quantum dilogarithms
\be
(q;q)_\infty (qX;q)_\infty \frac{1}{(X;q)_\infty} .
\ee
Once more, the KS theorem applies.
\end{proof}

We conclude by noting that the admissible series in $X^{-1}$ are remarkably simpler than the corresponding $X$-series in the factorisations in equations~\eqref{fact-admissible-2},~\eqref{fact-admissible-4}, and~\eqref{fact-admissible-5}. Particularly, they appear within products of $q$-theta functions in both examples of the $\FE$- and $\TT$-knots.

\section{The wave function from the state integral} \label{sec: stateintegral}
In this section, we consider the Andersen--Kashaev invariant $\chi_{\CK}(u;\hbar)$ in its original form as a~finite-dimensional state integral~\cite{ak, ak2, dimofte-rev}, whose integrand is a product of Faddeev’s quantum dilogarithm functions~\cite{faddeev}. We evaluate it directly at rational values of $\hbar$ by applying the techniques of~\cite{garou-kas}, thus providing a third method for the computation of the exact wave function. For simplicity, we focus on the case of the figure-eight knot and show that this third approach agrees with the two previous methods presented in Section~\ref{sec: main}.
A similar calculation in the case~${\mb^2=1}$ was done in unpublished work by Szabolcs Zakany.

\subsection{Evaluation at rational points}
Following the conventions of~\cite{ak}, the state integral for the $\FE$-knot takes the form\footnote{Note that the expressions in~\eqref{chi41} and~\eqref{41ak} differ by the global exponential factor $\re^{-\frac{\pi \ri}{12} (\mb^2+\mb^{-2})-\frac{x}{2}-\frac{x}{2 \mb^2} }$.}
\be \label{41ak}
\chi_{\FE}(u;\hbar) = \frac{1}{2 \pi \mb} \re^{-\frac{x}{2}-\frac{x}{2 \mb^2}-\frac{\ri x^2}{2 \pi \mb^2}} \int_{\IR + \ri 0} \frac{\Phi_{\mb}\left(\frac{x-y}{2 \pi \mb} \right)}{\Phi_{\mb}\left(\frac{y}{2 \pi \mb} \right)} \re^{\frac{\ri x y}{\pi \mb^2}}{\rm d}y ,
\ee
where $x = 2 \pi \mb u$ and $\hbar = 2 \pi \mb^2$, as before, while $\Phi_{\mb}(z)$ denotes the Faddeev's quantum dilogarithm. We refer to Appendix~\ref{app: faddeev} for its definition and some useful properties.
Let us define the function
\be
f(y) = \frac{\Phi_{\mb}\left(\frac{x-y}{2 \pi \mb} \right)}{\Phi_{\mb}\left(\frac{y}{2 \pi \mb} \right)} \re^{\frac{\ri x y}{\pi \mb^2}} ,
\ee
where we consider $x \in \IR$ fixed. The singularities of $f(y)$ are poles located at the points
\begin{subequations} \label{y-poles}
\begin{align}
&y= x - \pi \ri (2 M + 1 ) - \pi \ri \mb^2 (2 N + 1 ) , \\
&y= - \pi \ri (2 M + 1 ) - \pi \ri \mb^2 (2 N + 1 ) ,
\end{align}
\end{subequations}
where $M, N \in \IN$, as dictated by the singularities and the zeroes of Faddeev's quantum dilogarithm in Appendix~\ref{app: faddeev}.
Therefore, $f(y)$ is a holomorphic function in the upper half-plane.
Let us now fix the rational value
\be
\mb^2 = \frac{P}{Q} , \qquad P,Q \in \IZ_{>0} \ {\rm coprime} ,
\ee
and observe that the function $f(y)$ satisfies the relation
\begin{align}
f(y + 2 \pi \ri P) = f\bigl(y+2 \pi \ri \mb^2 Q\bigr) &= \Phi_{\mb}\left(\frac{x-y}{2 \pi \mb} - \ri \mb Q \right) \Phi_{\mb}\left(\frac{y}{2 \pi \mb} + \ri \mb Q \right)^{-1} \re^{\frac{\ri x y}{\pi \mb^2}- 2 x Q}\nonumber \\
&= f(y)\bigl( 1 + \re^{Q(x-y) - \pi \ri P Q} \bigr) \bigl( 1 + \re^{Q y + \pi \ri P Q} \bigr) \re^{-2 x Q} ,\label{f-shiftP}
\end{align}
and similarly
\begin{align}
f(y - 2 \pi \ri P)& = f\bigl(y-2 \pi \ri \mb^2 Q\bigr) = \Phi_{\mb}\left(\frac{x-y}{2 \pi \mb} + \ri \mb Q \right) \Phi_{\mb}\left(\frac{y}{2 \pi \mb} - \ri \mb Q \right)^{-1} \re^{\frac{\ri x y}{\pi \mb^2}+ 2 x Q} \nonumber\\
&= f(y) \bigl( 1 + \re^{Q(x-y) + \pi \ri P Q} \bigr)^{-1} \bigl( 1 + \re^{Q y - \pi \ri P Q} \bigr)^{-1} \re^{2 x Q} ,\label{f-shiftM}
\end{align}
where we have applied the periodicity formulae for $\Phi_{\mb}(z)$ at rational values in equations~\eqref{phibrat-perP} and~\eqref{phibrat-perM}.
Since $P,Q \in \IZ_{>0}$ and thus $\re^{\pi \ri P Q} = \re^{- \pi \ri P Q}$, it follows from equations~\eqref{f-shiftP} and~\eqref{f-shiftM} that $f(y)$ satisfies the functional equation
\be
f(y + 2 \pi \ri P) f(y - 2 \pi \ri P) = f(y)^2 .
\ee
It is straightforward to verify that
\be
f(y) (f(y)-f(y+2 \pi \ri P)) \ne 0 , \qquad \forall y \in \IR + \ri 0 ,
\ee
and therefore we can use~\cite[Lemma~2.1]{garou-kas} to write
\be \label{int-lemma}
\int_{\IR + \ri 0} f(y){\rm d}y = \left( \int_{\IR + \ri 0} - \int_{\IR + \ri 0 + 2 \pi \ri P} \right) \frac{f(y)}{1- f(y+2 \pi \ri P)/f(y)} {\rm d}y .
\ee
Let us now compute the integral on the right-hand side of equation~\eqref{int-lemma} applying the residue theorem. We start by using equation~\eqref{f-shiftP} to express the integrand in the form
\be \label{integrand}
\frac{f(y)}{1- f(y+2 \pi \ri P)/f(y)} = \frac{f(y) \re^{2 x Q}}{G(x,y)} ,
\ee
where we have introduced the function
\be
G(x,y) = \re^{2 x Q} - \bigl( 1 + \re^{Q(x-y) - \pi \ri P Q} \bigr) \bigl( 1 + \re^{Q y + \pi \ri P Q} \bigr) .
\ee
Since the numerator in equation~\eqref{integrand} is holomorphic in the upper half of the complex plane, the relevant singularities of the integrand are the solutions to $G(x,y) = 0$ at fixed $x \in \IR$ that lie within the domain of integration. Specifically, these are
\be \label{y-sing}
y_{\pm, k}(x) = 2 \pi \ri \frac{k}{Q} - \pi \ri P - \frac{1}{Q} \log \left( -\frac{1 + \re^{x Q} - \re^{2 x Q} \pm \sqrt{\Delta\bigl(\re^{x Q}\bigr)}}{2 \re^{x Q}} \right) ,
\ee
where $k \in \IZ$ such that $0 \le k \le P Q -1$, the function $\Delta(X)$ is defined by
\be \label{Delta}
\Delta(X) = 1- 2 X - X^2 - 2 X^3 + X^4 ,
\ee
and the branch of the logarithm is chosen in such a way that $\operatorname{Im} (y_{\pm,0}(x)) \in (0, 2 \pi/Q]$ for each value of $x \in \IR$. Note that $\Delta(X)$ is the same function introduced in equation~\eqref{classicD}, which appears in the formula for the solutions to the classical $A$-polynomial of the figure-eight knot in equation~\eqref{classicP} and in the formula for the function $\varphi_Q(X)$ obtained via the matrix formalism of Section~\ref{sec: 3rd-order} in equation~\eqref{8varphiQ}. Besides, the functions $y_{\pm, k}(x)$ only depend on $x$ through~${\re^{x Q}=X^Q}$.
We denote $y_{\pm}(x) = y_{\pm, 0}(x)$ for simplicity, so that
\be
y_{\pm, k}(x) = y_{\pm}(x) + 2 \pi \ri \frac{k}{Q} , \qquad 0 \le k \le P Q -1 .
\ee
For \smash{$y_{\pm, k}^{(\e)}(x) =y_{\pm, k}(x) + \e$} and $\e \rightarrow 0^+$, we have the asymptotic expansions
\be
\frac{1}{G\bigl( x, y_{\pm, k}^{(\e)}(x) \bigr)} = \mp \frac{1}{\e Q \sqrt{\Delta\bigl(\re^{x Q}\bigr)}} + \CO \bigl(\e^0\bigr) , \qquad 0 \le k \le P Q -1 .
\ee
Note that the leading-order term in the expansion above does not depend on the integer $k$. The integral in equation~\eqref{int-lemma} can then be evaluated via the sum of residues of the integrand in equation~\eqref{integrand} at the singularities in equation~\eqref{y-sing}. We find in this way that
\begin{align}
\int_{\IR + \ri 0} f(y){\rm d}y &= 2 \pi \ri \sum_{k=0}^{P Q - 1} \underset{y(x) = y_{\pm, k}(x)}{\operatorname{Res}} \left( \frac{f(y) \re^{2 x Q}}{G(x,y)} \right) \nonumber\\
&= - \frac{2 \pi \ri \re^{2 x Q} }{Q \sqrt{\Delta\bigl(\re^{x Q}\bigr)}} \sum_{k=0}^{P Q - 1} \re^{-\frac{2 x k}{P}} \bigl( \re^{F_k(y_+(x))} - \re^{F_k(y_-(x))} \bigr) ,\label{int-sol}
\end{align}
where we have introduced the notation
\be \label{Fk}
F_k(y(x)) = \frac{\ri Q x y(x)}{\pi P} + \log \Phi_{\mb}\left( \frac{x - y(x)}{2 \pi \mb} - \ri \mb \frac{k}{P} \right) - \log \Phi_{\mb}\left( \frac{y(x)}{2 \pi \mb} + \ri \mb \frac{k}{P} \right) ,
\ee
for $0 \le k \le P Q -1$. Therefore, substituting equation~\eqref{int-sol} into equation~\eqref{41ak}, we obtain that the state-integral invariant of the figure-eight knot at rational values is given exactly by
\be \label{state-fact}
\chi_{\FE}(u;\hbar) = \frac{\re^{\frac{x^2 Q}{2 \pi \ri P}+x\left( 2 Q -\frac{1}{2}-\frac{Q}{2 P}\right)} }{\ri \mb Q \sqrt{\Delta\bigl(\re^{x Q}\bigr)}} \sum_{k=0}^{P Q - 1} \re^{-\frac{2 x k}{P}} \bigl( \re^{F_k(y_+(x))} - \re^{F_k(y_-(x))} \bigr) .
\ee
We conclude by noting that $\Phi_{\mb}(z)$ for $\mb^2 \in \IQ_{>0}$ can be expressed explicitly in terms of more elementary functions~\cite{garou-kas}. The relevant formula is written in equation~\eqref{phib-rat} and can be used to simplify the computational implementation of equation~\eqref{state-fact} for arbitrary $P, Q \in \IZ_{>0}$ coprime.

\begin{figure}[htb!]
\centering
 \includegraphics[width=0.47\textwidth]{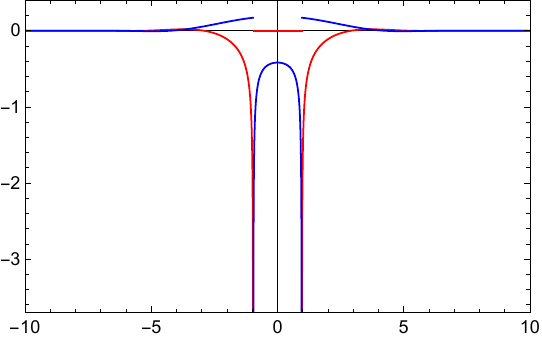} \quad \quad
\includegraphics[width=0.47\textwidth]{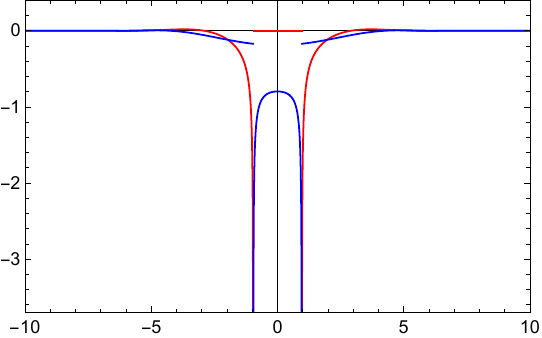} \\[2mm]
 \includegraphics[width=0.47\textwidth]{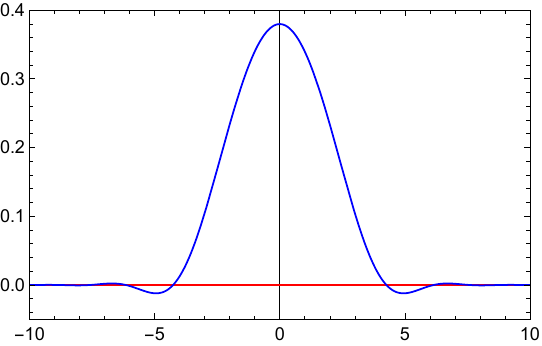}
 \caption{The state integral $\chi_{\FE}(u;\hbar)$ in the closed form of equation~\eqref{state-fact} as a function of $x \in \IR$, at the bottom, and its components \smash{$\chi_{\FE}^{(\pm)}(u;\hbar)$} in equation~\eqref{state-pieces}, in the top left and top right, respectively, for $P=Q=1$. The real and imaginary parts are displayed in blue and red.}
 \label{fig: state-unit}
\end{figure}
\begin{figure}[htb!]
\centering
 \includegraphics[width=0.47\textwidth]{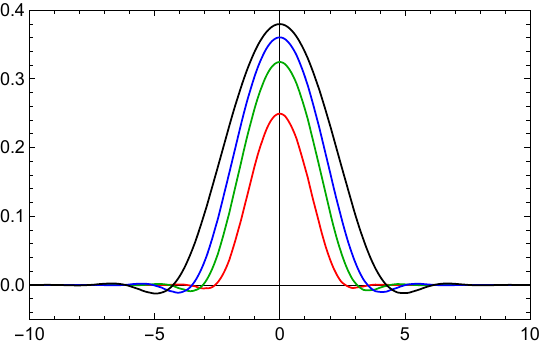} \quad \quad
 \includegraphics[width=0.47\textwidth]{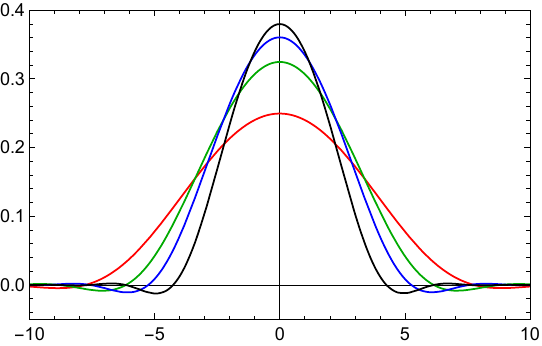}
 \caption{The real part of the state integral $\chi_{\FE}(u;\hbar)$ in the closed form of equation~\eqref{state-fact} as a function of $x \in \IR$ for rational values of $\mb^2$. We show $\mb^2 = 1/3$ (in red), $1/2$ (in green), $2/3$ (in blue), and $1$ (in black), on the left, and $\mb^2 = 3$ (in red), $2$ (in green), $3/2$ (in blue), and $1$ (in black), on the right. The imaginary part is identically zero and not displayed explicitly in the plot.}
 \label{fig: state-full}
\end{figure}

We will now comment on the most notable features of the exact expression for the non-perturbative wave function in equation~\eqref{state-fact}. It will be useful to introduce
\be \label{state-pieces}
 \chi_{\FE}^{(\pm)}(u;\hbar) = \frac{\re^{\frac{x^2 Q}{2 \pi \ri P}+x\left( 2 Q -\frac{1}{2}-\frac{Q}{2 P}\right)} }{\ri \mb Q \sqrt{\Delta\bigl(\re^{x Q}\bigr)}} \sum_{k=0}^{P Q - 1} \re^{-\frac{2 x k}{P}} \re^{F_k(y_{\pm}(x))} ,
\ee
where again $x \in \IR$, so that we write
\be \label{difference}
\chi_{\FE}(u;\hbar) = \chi_{\FE}^{(+)}(u;\hbar) - \chi_{\FE}^{(-)}(u;\hbar) .
\ee
Observe that the $S$-duality, which acts in the rational case via $P \leftrightarrow Q$ and $x \leftrightarrow x Q/P$, leaves~$\Delta\bigl(\re^{x Q}\bigr)$ unchanged while transforming $y_{\pm, k}(x)$ into $y_{\pm, k}(x) Q/P$, and therefore the closed formula for the state integral of the $\FE$-knot in equation~\eqref{state-fact} is $S$-invariant, as expected. Similarly, we also find that its components in equation~\eqref{state-pieces} are themselves invariant under $S$-duality.
Moreover, we have that
\be
\operatorname{Im}\bigl(\chi_{\FE}^{(+)}(u;\hbar)\bigr) = \operatorname{Im}\bigl( \chi_{\FE}^{(-)}(u;\hbar)\bigr) ,
\ee
while both $\operatorname{Re}\bigl( \chi_{\FE}^{(\pm)}(u;\hbar) \bigr)$ and $\operatorname{Im}\bigl( \chi_{\FE}^{(\pm)}(u;\hbar) \bigr)$ are even functions of $x \in \IR$ with singularities located at the points
\be
x = \frac{1}{Q} \log \left( \frac{3 \pm \sqrt{5}}{2} \right) ,
\ee
which correspond to the zeros of $\Delta\bigl(\re^{x Q}\bigr)$ in equation~\eqref{Delta} and, as such, do not depend on the value of $P$.
The discontinuities of \smash{$\operatorname{Re}\bigl( \chi_{\FE}^{(\pm)}(u;\hbar) \bigr)$} cancel with each other to give a smooth difference function in equation~\eqref{difference}, which is itself symmetric under $x \rightarrow -x$.
Remarkably, we find that the total state integral $\chi_{\FE}(u;\hbar)$ is a real-valued smooth square-integrable function of~${x \in \IR}$, while both functions \smash{$\chi_{\FE}^{(\pm)}(u;\hbar)$} are neither real-valued, smooth, or square-integrable when considered separately.
To illustrate these properties with an example, we show in Figure~\ref{fig: state-unit} the functions in equations~\eqref{state-fact} and~\eqref{state-pieces} for the benchmark choice $P=Q=1$. The same features and similar functional profiles are verified for various choices of $P, Q \in \IZ_{>0}$ coprime. We show in Figure~\ref{fig: state-full} the full non-perturbative wave function for a sample of small values of $\mb^2 \in \IQ_{>0}$.
We observe that $\chi_{\FE}(u;\hbar)$ has a Gaussian-type profile centered at $x=0$ with lateral oscillations around the $x$-axis that are quickly dampened.
The peak at the origin can be evaluated directly from equations~\eqref{y-sing},~\eqref{Fk}, and~\eqref{state-fact} for arbitrary $P, Q \in \IZ_{>0}$ coprime, yielding a result which is left unchanged by $P \leftrightarrow Q$, as expected from the $S$-invariance of the state integral and evident in Figure~\ref{fig: state-full}. In particular, the highest peak corresponds to the choice $P=Q=1$ and amounts to
\be
\chi_{\FE}(0; 2 \pi) = \frac{2}{\sqrt{3}} \sinh \left( \frac{V}{2 \pi} \right) , \qquad V = 2 \operatorname{Im} \bigl( \operatorname{Li}_2 \bigl( \re^{\frac{\pi \ri}{3}}\bigr)\bigr) = 2.0298832 \dots ,
\ee
where $\operatorname{Li}_2(z)$ is the standard dilogarithm and $V$ is the volume of the $\FE$-knot, as expected from~\cite{garou-kas}.
As $\mb^2<1$ decreases, the shape of the function $\chi_{\FE}(u;\hbar)$ tightens, while for increasing~${\mb^2>1}$, the shape widens, and in both cases the peak lowers.

We point out that the structure of the exact wave function at rational values for the figure-eight knot obtained here is tantalizingly close to what has been found in the spectral theory of quantum mirror curves. The case of $\hbar=2 \pi$, corresponding to $\mb^2=1$, was studied in detail in~\cite{mz-wv,mz-wv2}. As in equation~\eqref{difference}, the wave functions in~\cite{mz-wv,mz-wv2} are the sums of two individual functions corresponding to two different choices of branch cut in the mirror curve. Each component function is singular when considered alone. Still, the singularities cancel each other in the sum, and one obtains a square-integrable wave function in the end, similar to the one plotted in Figure~\ref{fig: state-unit}. We observe that, in the case of the quantum mirror curves studied in~\cite{mz-wv,mz-wv2}, one is working with exponentiated Heisenberg operators, and the resulting wave functions do not satisfy the nodal theorem for the Schr\"odinger operator typical of confining potentials, since the ground state changes sign in a small region. This same feature is also evident in our plots for the exact wave functions of the $\FE$-knot at rational points shown in Figures~\ref{fig: state-unit} and~\ref{fig: state-full}. Such a~property suggests that the wave function occurring in complex CS theory is a ground state for an appropriate spectral definition of the quantum $A$-polynomial. We will return to this point in the conclusions.

Finally, let us stress that the results above provide a third method for the computation of the exact wave function at rational values of $\hbar$ to be compared with the previous two presented in Section~\ref{sec: main}. We conclude by verifying numerically that our closed formula in equation~\eqref{state-fact} for the state integral of the figure-eight knot at rational values matches the results of Section~\ref{sec: 41}. Specifically, the two components \smash{$\chi_{\FE}^{(\pm)}(u;\hbar) $} in equation~\eqref{state-pieces} can be formally identified with the two terms in the holomorphic/antiholomorphic block decomposition in equation~\eqref{chi41}, which correspond to the conjugate and geometric classical branches, respectively. This identification will be clarified in the discussion in the following section.

\subsection{Factorization and two special subcases}
Let us apply the explicit formula for the Faddeev's quantum dilogarithm at the rational point $\mb^2=P/Q$, $P, Q \in \IZ_{>0}$ coprime, in equation~\eqref{phib-rat} to the expression for $F_k(y(x))$, $0 \le k \le PQ-1$, which is given in equation~\eqref{Fk}. After appropriately rearranging the terms, we obtain the factorization
\begin{gather} \label{Fk-fact}
\re^{F_k(y(x))} = \re^{F_0(y(x))} f_{0,k}(y(x)) f_{1,k}(y(x)) f_{2,k}(y(x)) ,
\end{gather}
where we have introduced the functions
\begin{subequations}
\begin{align}
f_{0,k}(y(x)) &= \big[\bigl( 1 - \re^{Q(x-y(x))+ \pi \ri (P+Q)} \bigr) \bigl( 1 - \re^{Q y(x)+ \pi \ri (P+Q)} \bigr)\big]^{\frac{k}{P Q}} , \\
f_{1,k}(y(x)) &=\prod_{j=1}^{Q-1}
\left[ \frac{\bigl(1 + \re^{x-y(x)+ \frac{\pi \ri P}{Q}(2j+1)} \bigr) \bigl(1 + \re^{y(x)+ \frac{\pi \ri P}{Q}(2j+1+2k/P)} \bigr)}{\bigl(1 + \re^{x-y(x)+ \frac{\pi \ri P}{Q}(2j+1-2k/P)} \bigr) \bigl(1 + \re^{y(x)+ \frac{\pi \ri P}{Q}(2j+1)} \bigr)} \right]^{j \over Q} , \\
f_{2,k}(y(x)) &=\prod_{j=1}^{P-1}
\left[ \frac{\bigl(1 + \re^{\frac{Q}{P}(x-y(x))+ \frac{\pi \ri Q}{P}(2j+1)} \bigr) \bigl(1 + \re^{\frac{Q}{P} y(x)+ \frac{\pi \ri Q}{P}(2j+1+2k/Q)} \bigr)}{\bigl(1 + \re^{\frac{Q}{P}(x-y(x))+ \frac{\pi \ri Q}{P}(2j+1-2k/Q)} \bigr) \bigl(1 + \re^{\frac{Q}{P} y(x)+ \frac{\pi \ri Q}{P}(2j+1)} \bigr)} \right]^{j \over P} ,
\end{align}
\end{subequations}
for $0 \le k \le PQ-1$. Thus, we can write the state integral for arbitrary rational values of $\mb^2$ in equation~\eqref{state-fact} in the equivalent form
\be \label{state-fact-PQ}
\chi_{\FE}(u;\hbar) = \frac{\re^{\frac{x^2 Q}{2 \pi \ri P}+x( 2 Q -\frac{1}{2}-\frac{Q}{2 P})} }{\ri \mb Q \sqrt{\Delta\bigl(\re^{x Q}\bigr)}} \big[\re^{F_0(y_+(x))} \CA(y_+(x)) - \re^{F_0(y_-(x))} \CA(y_-(x)) \big] ,
\ee
where the function $\CA(y(x))$ is defined by
\be
\CA(y(x)) = \sum_{k=0}^{P Q - 1} \re^{-\frac{2 x k}{P}} f_{0,k}(y(x)) f_{1,k}(y(x)) f_{2,k}(y(x)) ,
\ee
and $F_0(y(x))$ can be expressed explicitly in terms of more elementary functions using equation~\eqref{phib-rat}.
We stress that the formula for the state integral in equation~\eqref{state-fact-PQ} allows us to more clearly separate the contributions from the two non-abelian flat $\mathrm{SL}(2,\IC)$-connections of the $\FE$-knot labeled by $\alpha = {\rm geom}, \mathrm{conj}$, as in the holomorphic/antiholomorphic block decomposition in equation~\eqref{exactconj}. Indeed, expanding the functions $F_0(y_{\mp}(x))$ in powers of $X, X^{-1}$ and suitably identifying the terms in the expansions, we derive the integer constants \smash{$c^{(\alpha)}_{\pm s}$}, $s \in \IZ_{>0}$, previously written in equations~\eqref{cs41geom} and~\eqref{cs41conj}, for the geometric and conjugate branches, respectively.

Let us consider now the two special choices $\mb^2 = P$ and $\mb^2 = 1/Q$ for $P, Q \in \IZ_{>0}$. In both subcases, the closed formula for the state integral of the figure-eight knot in equation~\eqref{state-fact-PQ} specializes in a correspondingly simpler and more compact form.
We start by choosing $\mb^2= P$ positive integer. After some manipulations, the factorization formula in equation~\eqref{Fk-fact} reduces to the relation
\be \label{Fk-P-fact}
\re^{F_k(y(x))} = \re^{F_0(y(x))} \prod_{j=0}^{k-1} \bigl(1 + \re^{\frac{1}{P}(x-y(x))-\frac{\pi \ri}{P} (2j+1)} \bigr) \bigl(1 + \re^{\frac{1}{P}y(x)+\frac{\pi \ri}{P} (2j+1)} \bigr) ,
\ee
where $0 \le k \le P -1$, and we can write the state integral in equation~\eqref{state-fact-PQ} for $\mb^2=P$ as
\be \label{state-fact-P}
\chi_{\FE}(u;\hbar) = \frac{\re^{\frac{x^2}{2 \pi \ri P}+x( \frac{3}{2}-\frac{1}{2 P})} }{\ri \mb \sqrt{\Delta(\re^x)}} \big[ \re^{F_0(y_+(x))} \CB(y_+(x)) - \re^{F_0(y_-(x))} \CB(y_-(x)) \big] ,
\ee
where the function $\CB(y(x))$ is simply defined by
\be
\CB(y(x)) = \sum_{k=0}^{P - 1} \re^{-\frac{2 x k}{P}} \prod_{j=0}^{k-1} \bigl(1 + \re^{\frac{1}{P}(x-y(x))-\frac{\pi \ri}{P} (2j+1)} \bigr) \bigl(1 + \re^{\frac{1}{P}y(x)+\frac{\pi \ri}{P} (2j+1)} \bigr) ,
\ee
and $F_0(y(x))$ can be expressed explicitly by means of equation~\eqref{phib-ratP}.
Let us now choose $\mb^{-2}= Q$ positive integer. After some manipulations, the factorization formula in equation~\eqref{Fk-fact} becomes
\be \label{Fk-M-fact}
\re^{F_k(y(x))} = \re^{F_0(y(x))} \prod_{j=0}^{k-1} \bigl(1 + \re^{x-y(x)-\frac{\pi \ri}{Q} (2j+1)} \bigr) \bigl(1 + \re^{y(x)+\frac{\pi \ri}{Q} (2j+1)} \bigr) ,
\ee
and the exact partition function in equation~\eqref{state-fact-PQ} for $\mb^2=1/Q$ assumes the simplified form
\be \label{state-fact-M}
\chi_{\FE}(u;\hbar) = \frac{\re^{\frac{x^2 Q}{2 \pi \ri}+x( \frac{3 Q}{2} -\frac{1}{2} )} }{\ri \mb Q \sqrt{\Delta\bigl(\re^{x Q}\bigr)}} \big[ \re^{F_0(y_+(x))} \CC(y_+(x)) - \re^{F_0(y_-(x))} \CC(y_-(x)) \big] ,
\ee
where the function $\CC(y(x))$ is given by
\be
\CC(y(x)) = \sum_{k=0}^{Q - 1} \re^{-2 x k} \prod_{j=0}^{k-1} \bigl(1 + \re^{x-y(x)-\frac{\pi \ri}{Q} (2j+1)} \bigr) \bigl(1 + \re^{y(x)+\frac{\pi \ri}{Q} (2j+1)} \bigr) ,
\ee
and $F_0(y(x))$ can be expressed explicitly by means of equation~\eqref{phib-ratQ}.
We observe that the reduced factorization formulae in equations~\eqref{Fk-P-fact} and~\eqref{Fk-M-fact} for $\mb^2 = P$ and $\mb^2 = 1/Q$ with~${P, Q \in \IZ_{>0}}$ can alternatively be derived by substituting $\ri \mb k / P = \ri \mb^{-1} k$ and $\ri \mb k / P = \ri \mb k$ into the arguments of the Faddeev's quantum dilogarithms in equation~\eqref{Fk} and applying the periodicity properties in equations~\eqref{PM} and~\eqref{MM} and equations~\eqref{PP} and~\eqref{MP} to the resulting formulae for $F_k(y(x))$, $0 \le k \le PQ-1$, respectively. In the general case of $\mb^2 = P/Q$, instead, we cannot resort to these quasi-periodicity properties because the functions $F_k(y(x))$ contain, in general, a fractional shift in the arguments of the Faddeev's quantum dilogarithms.

\section{Conclusions}\label{sec-conclusions}
In this paper, we have made various observations on the structure of the wave functions in CS theory with gauge group $\mathrm{SL}(2,\IC)$ on the complement of a hyperbolic knot in the three-sphere.
We have first conjectured explicit integrality properties for the resummed WKB expansion of the wave function, which can be tested directly using the decomposition into holomorphic/antiholomorphic blocks.
Our integrality conjecture guarantees that the singularities appearing in the holomorphic blocks at rational values of $\hbar$ get canceled against those occurring in the corresponding antiholomorphic blocks.
We have then developed various techniques to compute the wave function in the rational case effectively. In particular, we have analyzed the quantum $A$-polynomial at rational points by exploiting the underlying quasi-periodic structure and solved the associated $q$-difference equation in closed form in the general cases of orders two and three.
The calculation methods we have introduced have been applied to the examples of the figure-eight and three-twist knots, and our conjectural statements have been verified.
Finally, we have performed a direct evaluation of the Andersen--Kashaev state integral of the figure-eight knot at rational values, which gave us further insights into the properties of the exact wave functions. All computational approaches yielded compatible results.

Our investigation raises two main questions.
The first one concerns the interpretation of the integrality structure of the resummed WKB expansion of the wave function in terms of an enumerative problem, or a BPS counting problem, in the dual supersymmetric theory~\cite{dgg}, either directly or potentially exploiting results from the resurgent analysis of the perturbative wave functions performed in~\cite{ggm2}.
The second one concerns the shape of the wave function obtained by directly evaluating the state integral. As highlighted in Section~\ref{sec: stateintegral}, this is enticingly similar to the functional profiles of the ground state wave functions computed in the context of the TS/ST correspondence~\cite{mz-wv, mz-wv2}. This observation raises the issue of understanding the AJ conjecture of~\cite{stavros-aj} from a purely quantum-mechanical point of view.
The AJ conjecture can be explained, {\it a priori}, as a consequence of the quantization of the classical moduli space described by the $A$-polynomial, as pointed out in~\cite{gukov} and briefly recalled in Section~\ref{sec: theory}.
However, the precise details of this quantization procedure still need to be fully understood from a physical perspective. We know that the quantization is non-trivial and cannot be obtained by simply promoting the classical variables $X$, $P$ to Weyl operators. In addition, the classical phase space is not compact, which carries various problems concerning its appropriate quantization. From a quantum-mechanical angle, in such a situation, we would generically expect to have, at most, metastable states with complex energies. We are, therefore, tempted to conjecture that the observed non-trivial quantization of the $A$-polynomial is required to obtain a normalizable wave function with precisely zero energy. We hope to address this question in the near future.

\appendix

\section[Computing the epsilon-expansions]{Computing the $\boldsymbol{\epsilon}$-expansions} \label{app: eps}
For completeness, we include here the details of the computations performed in Section~\ref{sec: cancellation}.
We observe that, as a consequence of equations~\eqref{ahatf-def} and~\eqref{struc-an}, the coefficient functions $\hat{a}_{\pm n}(q)$ satisfy the relation
 \be
 \hat{a}_{\pm n}(q) q^{-\sigma_{\pm n}}= \sum_{k|n} k D_{\pm k} \bigl( q^{n \over k} \bigr) q^{-{n \over k} \sigma_{\pm k}} \sum_{j=0}^{k-1} q^{{n \over k}j} , \qquad n \in \IZ_{>0} ,
 \ee
and we write the $\IZ$-polynomials $D_{\pm k} (q)$ explicitly as
\be
D_{\pm k}(q)= \sum_{\ell=0}^{\deg{D_{\pm k}}} D_{\pm k, \ell} q^\ell , \qquad k \in \IZ_{>0} ,
\ee
where $D_{\pm k, \ell}$ is an integer and $\deg{D_{\pm k}}$ denotes the degree of the polynomial.
Let us now take
\be
q=\re^{2 \pi \ri {P \over Q}} , \qquad P,Q \in \IZ_{>0} \ {\rm coprime} ,
\ee
while we introduce
\be
q_{\e}= \re^{\ri \hbar_{\e}} , \qquad \hbar_{\e}= 2 \pi {P \over Q}+ \epsilon , \qquad 0 < \e \ll 1 ,
\ee
and consider the limit $\e \rightarrow 0$.
In the case $n= s Q$, $s \in \IZ_{>0}$, we have the $\e$-expansions
\begin{subequations}
\begin{gather}
D_{\pm k}\bigl( q_{\e}^{s Q \over k} \bigr) = \left( 1 + \ri \e q {\partial \over \partial q} \right) D_{\pm k}\bigl( q^{s Q \over k} \bigr) + \CO\bigl(\e^2\bigr) , \\
\sum_{j=0}^{k-1} q_{\e}^{{s Q \over k}(j-\sigma_{\pm k})} = \left( 1 + \ri \e q {\partial \over \partial q} \right) \sum_{j=0}^{k-1} q^{{s Q \over k}(j-\sigma_{\pm k})} + \CO\bigl(\e^2\bigr) ,
\end{gather}
\end{subequations}
and therefore also
\be \label{exp-ahat}
 \hat{a}_{\pm s Q}(q_{\e}) q_{\e}^{-\sigma_{\pm s Q}} = \left( 1 + \ri \e q {\partial \over \partial q} \right) \hat{a}_{\pm s Q}(q) q^{-\sigma_{\pm s Q}} + \CO\bigl(\e^2\bigr) .
\ee
We now observe that
\be \label{exp-frac}
\frac{1}{q_{\e}^{s Q }-1} = -\frac{\ri}{s Q \e}-\frac{1}{2} + \CO(\e) ,
\ee
and, combining equations~\eqref{exp-ahat} and~\eqref{exp-frac}, we finally obtain the NLO $\e$-expansion in equation~\eqref{pole}.
On the dual side, we consider
\be
q_D=q^{Q^2 \over P^2}=\re^{2 \pi \ri {Q \over P}} , \qquad q_{\e, D}= \re^{4 \pi^2 \ri \over \hbar_{\e}} ,
\ee
and take again the limit $\e \rightarrow 0$. It follows that
\be
q_{\e, D}= q_D \re^{- \ri \e {Q^2 \over P^2}+ \CO(\e^2)} .
\ee
Similarly to the case above, when $n= s P$, $s \in \IZ_{>0}$, we have the $\e$-expansions
\begin{subequations}
\begin{gather}
D_{\pm k}\Big( q_{\e,D}^{s P \over k} \Big) = \left( 1 - \ri \e \frac{Q^2}{P^2} q_D {\partial \over \partial q_D} \right) D_{\pm k}\Big( q_D^{s P \over k} \Big) + \CO\bigl(\e^2\bigr) , \\
\sum_{j=0}^{k-1} q_{\e,D}^{{s P \over k}(j-\sigma_{\pm k})} = \left( 1 - \ri \e \frac{Q^2}{P^2} q_D {\partial \over \partial q_D} \right) \sum_{j=0}^{k-1} q_D^{{s P \over k}(j-\sigma_{\pm k})} + \CO\bigl(\e^2\bigr) ,
\end{gather}
\end{subequations}
and therefore also
\be \label{exp-ahatD}
 \hat{a}_{\pm s P}(q_{\e,D}) q_{\e,D}^{-\sigma_{\pm s P}} = \left( 1 - \ri \e \frac{Q^2}{P^2} q_D {\partial \over \partial q_D} \right) \hat{a}_{\pm s P}(q_{D}) q_{D}^{-\sigma_{\pm s P}} + \CO\bigl(\e^2\bigr) .
\ee
Finally, we note that
\be \label{exp-fracD}
\frac{1}{q_{\e,D}^{s P}-1} = \frac{\ri P}{s Q^2 \e}-\frac{1}{2} + \frac{\ri}{2 \pi s Q} + \CO(\e) ,
\ee
and, combining equations~\eqref{exp-ahatD} and~\eqref{exp-fracD}, we obtain the NLO $\e$-expansion in equation~\eqref{poleD}.

\section{An alternative way of obtaining the integer sequences} \label{app: rational2}
We present an alternative, simple way of computing the integer sequences $\{c_{\pm s} \}$, $s \in \IZ_{>0}$, defined in equations~\eqref{cs-sequenceP} and~\eqref{cs-sequenceM}, using the results on the exact wave function at rational values obtained in Section~\ref{sec: rational1}. Let us denote by \smash{$\phi_{(\pm)}^{\rm WKB} (X; q)$} the contributions to the holo\-morphic component \smash{$\phi^{\rm WKB} (X; q)$} from the positive and negative powers of $X$, respectively, so that equation~\eqref{phi-struc} assumes the form
\be
\phi^{\rm WKB} (X; q) = \phi_{(+)}^{\rm WKB} (X; q) + \phi_{(-)}^{\rm WKB} (X; q) ,
\ee
and analogously for the antiholomorphic component $\phi^{\rm WKB} (X_D; q_D)$.
We fix $\hbar$ as in equation~\eqref{rathbar} and consider the formula for $\phi^{\rm WKB} (X; q)+\phi^{\rm WKB} (X_D; q_D)$ in equation~\eqref{phirat-final}.
Setting the values $P=Q=1$, the $X$-series becomes
\be \label{intP1}
\frac{\ri}{2 \pi} \sum_{s \ge 1} \frac{X^s}{s^2} \left[( 2 \pi \ri s + 1 - s \log X + 4 \pi \ri \sigma_s ) \hat{a}_s(1) - 4 \pi \ri \frac{\d}{\d q} \hat{a}_s(q)_{q=1} \right] .
\ee
Let us now go back to the sum in equation~\eqref{phi-sum} and apply the change of variable $\hbar \rightarrow -\hbar$, which transforms $q$, $q_D$ into $q^{-1}$, $q_D^{-1}$, while keeping $X$, $X_D$ fixed.
After performing the $\e$-expansion as in Section~\ref{sec: cancellation} and then taking $P=Q=1$ in the sum $\phi_{(+)}^{\rm WKB} \bigl(X; q^{-1}\bigr) + \phi_{(+)}^{\rm WKB} \bigl(X_D; q_D^{-1}\bigr)$, we find the very close but not identical quantity
\be \label{intP2}
\frac{\ri}{2 \pi} \sum_{s \ge 1} \frac{X^s}{s^2} \left[( 2 \pi \ri s - 1 + s \log X + 4 \pi \ri \sigma_s) \hat{a}_s(1) - 4 \pi \ri \frac{\d}{\d q} \hat{a}_s(q)_{q=1} \right] .
\ee
By taking the difference of equations~\eqref{intP1} and~\eqref{intP2}, we arrive at
\be \label{intP-final}
\frac{\ri}{\pi} \sum_{s \ge 1} \frac{X^s}{s^2} (1-s \log X) \hat{a}_s(1) .
\ee
Therefore, if we compute the function
\be
\big[ \bigl(\phi_{(+)}^{\rm WKB} (X; q) + \phi_{(+)}^{\rm WKB} (X_D; q_D) \bigr) - \bigl( \phi_{(+)}^{\rm WKB} \bigl(X; q^{-1}\bigr) + \phi_{(+)}^{\rm WKB} \bigl(X_D; q_D^{-1}\bigr) \bigr) \big]_{P=Q=1} ,
\ee
its expansion in powers of $X$ is precisely the series in equation~\eqref{intP-final}. From the coefficients of this expansion, we can easily extract the desired integers $c_s = \hat{a}_s(1)$ for $s \in \IZ_{>0}$.
Let us now consider the $X^{-1}$-series in equation~\eqref{phirat-final}. Again, setting the values $P=Q=1$, we find
\be \label{intM1}
\frac{\ri}{2 \pi} \sum_{s \ge 1} \frac{X^{-s}}{s^2} \left[( 2 \pi \ri s + 1 + s \log X) \hat{a}_{-s}(1) - 4 \pi \ri \frac{\d}{\d q} \hat{a}_{-s}(q)_{q=1} \right] .
\ee
As before, we go back to the sum in equation~\eqref{phi-sum} and apply the change of variable $\hbar \rightarrow -\hbar$, which transforms $q$, $q_D$ into $q^{-1}$, $ q_D^{-1}$, while keeping $X$, $X_D$ fixed.
After performing the $\e$-expansion as in Section~\ref{sec: cancellation} and then taking $P=Q=1$ in the sum
\[\phi_{(-)}^{\rm WKB} \bigl(X; q^{-1}\bigr) + \phi_{(-)}^{\rm WKB} \bigl(X_D; q_D^{-1}\bigr),
\]
 we find the slightly different quantity
\be \label{intM2}
\frac{\ri}{2 \pi} \sum_{s \ge 1} \frac{X^{-s}}{s^2} \left[( 2 \pi \ri s - 1 - s \log X) \hat{a}_{-s}(1) - 4 \pi \ri \frac{\d}{\d q} \hat{a}_{-s}(q)_{q=1} \right] .
\ee
By taking the difference of equations~\eqref{intM1} and~\eqref{intM2}, we arrive at
\be \label{intM-final}
\frac{\ri}{\pi} \sum_{s \ge 1} \frac{X^{-s}}{s^2} (1+s \log X) \hat{a}_{-s}(1) .
\ee
Accordingly, if we compute the function
\be
\big[ \bigl(\phi_{(-)}^{\rm WKB} (X; q) + \phi_{(-)}^{\rm WKB} (X_D; q_D) \bigr) - \bigl( \phi_{(-)}^{\rm WKB} \bigl(X; q^{-1}\bigr) + \phi_{(-)}^{\rm WKB} \bigl(X_D; q_D^{-1}\bigr) \bigr) \big]_{P=Q=1} ,
\ee
its expansion in powers of $X^{-1}$ is precisely the series in equation~\eqref{intM-final}. From the coefficients of this expansion, we can straightforwardly obtain the desired integers $c_{-s} = \hat{a}_{-s}(1)$ for $s \in \IZ_{>0}$.

\section{Faddeev's quantum dilogarithm and other special functions} \label{app: faddeev}
The quantum dilogarithm is the function of two variables defined by the series~\cite{fk, Kirillov}
\be \label{eq: dilog}
(z q^{\alpha}; q)_{\infty} = \prod_{i=0}^{\infty} \bigl(1- z q^{\alpha+i}\bigr) , \qquad \alpha \in \IR ,
\ee
which is analytic in $z,q \in \IC$ with $|q| <1$ and admits asymptotic expansions around $q$ a root of unity.
The $q$-Pochhammer symbols, also known as $q$-shifted factorials, are defined by
\begin{subequations} \label{eq: qFactor}
\begin{gather}
(z; q)_n = \prod_{i=0}^{n-1} \bigl(1- z q^i\bigr) , \qquad n \in \IZ_{>0} , \label{eq: qFactorP}\\
(z; q)_{-n} = \frac{1}{(z q^{-n}; q)_n} = \frac{(-z)^{-n} q^{n(n+1)/2}}{\bigl(z^{-1} q; q\bigr)_n} = \frac{1}{\bigl(z q^{-1}; q^{-1}\bigr)_n} , \qquad n \in \IZ_{>0} , \label{eq: qFactorM}
\end{gather}
\end{subequations}
with $(z; q)_0=1$. Equivalently, we can write
\be \label{eq: qFactor2}
(z; q)_n = \frac{(z; q)_{\infty}}{(z q^n; q)_{\infty}} , \qquad n \in \IZ .
\ee
Moreover, the quantum dilogarithm satisfies the $q$-binomial theorem, which can be expressed as
\begin{gather} \label{q-binomial}
\frac{(x z; q)_{\infty}}{(z ; q)_{\infty}} = \sum_{n=0}^\infty \frac{(x; q)_n}{(q ; q)_n} z^n ,
\end{gather}
where $x, z, q \in \IC$ and $|q| <1$, and implies the following $q$-series identities:
\begin{subequations} \label{eq: qID}
\begin{gather}
(q z; q)_{\infty} = \sum_{n=0}^\infty \frac{(-1)^n q^{n(n+1)/2}}{(q ; q)_n} z^n = \frac{1}{\bigl(z ; q^{-1}\bigr)_{\infty}} , \label{eq: qID-1}\\
\frac{1}{(z ; q)_{\infty}} = \sum_{n=0}^\infty \frac{z^n}{(q ; q)_n} = \bigl(q^{-1} z; q^{-1}\bigr)_{\infty} . \label{eq: qID-2}
\end{gather}
\end{subequations}

The Faddeev's quantum dilogarithm $\Phi_{\mb}(z)$ is defined in the strip $| \operatorname{Im} (z) | < | \operatorname{Im} (c_{\mb}) |$, where
\be
c_{\mb} = \ri \bigl(\mb + \mb^{-1}\bigr)/2 ,
\ee
by the integral representation~\cite{faddeev, fk}
\be \label{eq: intPhib}
\Phi_{\mb}(z) = \exp \left( \int_{\IR + \ri \epsilon} \frac{\re^{-2 \ri z y}}{4 \sinh(y \mb ) \sinh\bigl(y \mb^{-1}\bigr)} \frac{\rd y}{y} \right) ,
\ee
which implies the symmetry properties
\be \label{eq: symmPhib}
\Phi_{\mb}(z) = \Phi_{-\mb}(z) = \Phi_{\mb^{-1}}(z) .
\ee
When $\operatorname{Im}\big(\mb^2\big) > 0$, the formula in equation~\eqref{eq: intPhib} is equivalent to
\be \label{eq: seriesPhib}
\Phi_{\mb}(z) = \frac{\big( \re^{2 \pi \mb (z + c_{\mb})}; q\big)_{\infty}}{\big( \re^{2 \pi \mb^{-1} (z - c_{\mb})}; \tilde{q}\big)_{\infty}} = \prod_{n=0}^{\infty} \frac{1- \re^{2 \pi \mb (z + c_{\mb})} q^n}{1- \re^{2 \pi \mb^{-1} (z - c_{\mb})} \tilde{q}^n} ,
\ee
where
\be
q = \re^{2 \pi \mathrm{i} \mb^2} , \qquad \tilde{q} = \re^{- 2 \pi \mathrm{i} \mb^{-2}} .
\ee
Note that the function in equation~\eqref{eq: seriesPhib} can be extended to the region $\operatorname{Im}\bigl(\mb^2\bigr) < 0$ by means of equation~\eqref{eq: symmPhib} and further admits an analytic continuation to all values of $\mb$ such that $\mb^2 \notin \IR_{\le 0}$.
Moreover, $\Phi_{\mb}(z)$ can be extended to the whole complex $z$-plane as a meromorphic function with an essential singularity at infinity, poles at the points
\be
z = c_{\mb} + \ri m \mb + \ri n \mb^{-1} ,
\ee
and zeros at the points
\be
z = -c_{\mb} - \ri m \mb - \ri n \mb^{-1} ,
\ee
for $m,n \in \IN$. It satisfies the inversion formula
\be
\Phi_{\mb}(z) \Phi_{\mb}(-z) = \re^{\pi \ri z^2} \Phi_{\mb}(0)^2 , \qquad \Phi_{\mb}(0) = \left( \frac{q}{\tilde{q}} \right)^{1/48} = \re^{\pi \ri (\mb^2 + \mb^{-2})/24} ,
\ee
the complex conjugation formula
\be
\overline{\Phi_{\mb}(z)} = \frac{1}{\Phi_{\overline{\mb}}(\overline{z}) } ,
\ee
and is a quasi-periodic function. Precisely, as a direct consequence of its definition in equation~\eqref{eq: seriesPhib}, we find the relations
\begin{subequations} \label{periodicity}
\begin{gather}
\Phi_{\mb}(z + \ri \mb k) = \Phi_{\mb}(z) \prod_{j=0}^{k-1} \bigl( 1 + \re^{2 \pi \mb z + \pi \ri \mb^2 (2 j +1)} \bigr)^{-1} , \label{PP}\\
\Phi_{\mb}(z - \ri \mb k) = \Phi_{\mb}(z) \prod_{j=0}^{k-1} \bigl( 1 + \re^{2 \pi \mb z - \pi \ri \mb^2 (2 j +1)} \bigr) , \label{MP} \\
\Phi_{\mb}\bigl(z + \ri \mb^{-1} k\bigr) = \Phi_{\mb}(z) \prod_{j=0}^{k-1} \bigl( 1 + \re^{2 \pi \mb^{-1} z + \pi \ri \mb^{-2} (2 j +1)} \bigr)^{-1} , \label{PM} \\
\Phi_{\mb}\bigl(z - \ri \mb^{-1} k\bigr) = \Phi_{\mb}(z) \prod_{j=0}^{k-1} \bigl( 1 + \re^{2 \pi \mb^{-1} z - \pi \ri \mb^{-2} (2 j +1)} \bigr) , \label{MM}
\end{gather}
\end{subequations}
where $k \in \IZ_{>0}$. If we consider the special case of $\mb^2 \in \IQ_{>0}$ and, specifically, take
\be \label{rat-b2}
\mb^2 = \frac{P}{Q} , \qquad P,Q \in \IZ_{>0} \ {\rm coprime} ,
\ee
then the periodicity formulae above give in particular
\begin{subequations}
\begin{align}
\Phi_{\mb}(z + \ri \mb Q) &= \Phi_{\mb}(z) \bigl( 1 + \re^{2 \pi \mb Q z + \pi \ri P Q} \bigr)^{-1} , \label{phibrat-perP} \\
\Phi_{\mb}(z - \ri \mb Q) &= \Phi_{\mb}(z) \bigl( 1 + \re^{2 \pi \mb Q z - \pi \ri P Q} \bigr)^ . \label{phibrat-perM}
\end{align}
\end{subequations}
Finally, in the above rational case of equation~\eqref{rat-b2}, it was shown in~\cite{garou-kas} that Faddeev's quantum dilogarithm can be expressed as
\begin{align}
\Phi_{\mb} \left( \frac{z}{2 \pi \mb} \right)={}& \exp\left[\frac{\ri}{2 \pi P Q} \right. \operatorname{Li}_2\bigl( \re^{w(z)} \bigr) + \left( 1 + \frac{\ri w(z)}{2 \pi P Q} \right) \log \bigl( 1 - \re^{w(z)}\bigr) \nonumber\\
&\left. - \log D_Q\bigl( \re^{\frac{w(z)}{Q}}; \re^{2 \pi \ri \frac{P}{Q}} \bigr) - \log D_P\bigl( \re^{\frac{w(z)}{P}}; \re^{2 \pi \ri \frac{Q}{P}} \bigr) \right] ,\label{phib-rat}
\end{align}
where $w(z) = Q z + \pi \ri (P + Q)$, $\operatorname{Li}_2(z)$ is the standard dilogarithm, and
\be
D_N(z; q)= \prod_{j=1}^{N-1} \bigl(1- z q^j \bigr)^{j/N} , \qquad N \in \IZ_{>0} ,
\ee
is the $N$-th cyclic quantum dilogarithm with $D_1(z; q) = 1$. In the special subcases of $\mb^2 = P$ positive integer and $\mb^2 = 1/Q$ positive unit fraction, the formula in equation~\eqref{phib-rat} simplifies into
\begin{subequations}
\begin{align}
\Phi_{\mb} \left( \frac{z}{2 \pi \mb} \right) ={}& \exp\left[\frac{\ri}{2 \pi P} \operatorname{Li}_2\bigl( - \re^{z + \pi \ri P} \bigr) + \left(\frac{P-1}{2 P} + \frac{\ri z}{2 \pi P} \right) \log \bigl( 1 + \re^{z + \pi \ri P}\bigr) \right. \nonumber\\
&\left. -\log D_P\bigl( -\re^{\frac{z}{P} + \frac{\pi \ri}{P}}; \re^{\frac{2 \pi \ri}{P}} \bigr) \right] , \label{phib-ratP} \\
\Phi_{\mb} \left( \frac{z}{2 \pi \mb} \right) ={}& \exp\left[\frac{\ri}{2 \pi Q} \operatorname{Li}_2\bigl( - \re^{Q z + \pi \ri Q} \bigr) + \left(\frac{Q-1}{2 Q} + \frac{\ri z}{2 \pi} \right) \log \bigl( 1 + \re^{Q z + \pi \ri Q}\bigr) \right.\nonumber\\
&\left. -\log D_Q\bigl( -\re^{z + \frac{\pi \ri}{Q}}; \re^{\frac{2 \pi \ri}{Q}} \bigr) \right] , \label{phib-ratQ}
\end{align}
\end{subequations}
respectively.

\subsection*{Acknowledgements}
We thank Stavros Garoufalidis, Jie Gu, and Sergei Gukov for valuable discussions. We are particularly grateful to Szabolcs Zakany for an initial collaboration on this subject some years ago. Finally, we thank the anonymous referees for providing detailed reports that were instrumental in improving the paper.
This research is supported in part by the ERC-SyG project ``Recursive and Exact New Quantum Theory'' (ReNewQuantum), which received funding from the European Research Council (ERC) under the European Union's Horizon 2020 research and innovation program, grant agreement no.~810573.

\pdfbookmark[1]{References}{ref}
\LastPageEnding

\end{document}